\documentclass[amsmath,amssymb,aps,pre,12pt,longbibliography]{revtex4-1}

\usepackage{bm}
\usepackage{dcolumn}
\usepackage{graphicx}
\usepackage[caption=false]{subfig}

\begin{document}
\preprint{APS/123-QED}

\title{Dynamics of phase slips in systems with time-periodic modulation}
\author{Punit Gandhi}
 \email{punit\_gandhi@berkeley.edu}
 \affiliation{Department of Physics, University of California, Berkeley CA 94720, USA}
 
 \author{C\'edric Beaume}
 \email{ced.beaume@gmail.com}
\affiliation{Department of Aeronautics, Imperial College London, London SW7 2AZ, UK}

\author{Edgar Knobloch}
 \email{knobloch@berkeley.edu}
\affiliation{Department of Physics, University of California, Berkeley CA 94720, USA}

\date{\today}

\begin{abstract}
The Adler equation with time-periodic frequency modulation is studied. A series of 
resonances between the period of the frequency modulation and the time scale for the 
generation of a phase slip is identified. The resulting parameter space structure is 
determined using a combination of numerical continuation, time simulations and asymptotic 
methods. Regions with an integer number of phase slips per period are separated by regions 
with noninteger numbers of phase slips, and include canard trajectories that drift along 
unstable equilibria. Both high and low frequency modulation is considered. An adiabatic 
description of the low frequency modulation regime is found to be accurate over a large 
range of modulation periods. 

\end{abstract}

\pacs{Valid PACS appear here}

\maketitle

\section{Introduction}
\label{sec:intro}

This paper is devoted to a study of the nonautonomous Adler equation \cite{adler1946study}
\begin{equation}
\dot{\theta} = r(t) - \sin \theta.
\label{eq:adler}
\end{equation}
When $r$ is independent of time this equation describes phase synchronization between a pair of coupled oscillators. In this case $\theta\equiv\phi_1-\phi_2$ represents the difference in the phases $\phi_j$ of the two oscillators and $r$ represents the normalized frequency difference. When $|r|<1$ the equation describes a phase-locked state; when $|r|>1$ the phase difference increases or decreases monotonically, corresponding to repeated phase slips. The transition between these two states is an example of a SNIPER (saddle-node infinite period) or SNIC (saddle-node on an invariant circle) bifurcation \cite{strogatz2014nonlinear}. In this bifurcation the phase slip period diverges like $1/\sqrt{r-1}$ as $r$ decreases towards $r=1$, in contrast to transitions associated with global bifurcations.

The nonautomous equation (\ref{eq:adler}) with $r=r(t)$ and $r(t)$ a periodic function of time thus describes the effects of temporal modulation of the SNIPER bifurcation. Such modulation is of interest since for part of the modulation cycle the oscillators may be phase-locked while for the rest of the cycle they may undergo phase slips. In this paper we show that the interplay between these two states is complex, and characterize the resulting behavior for both high and low frequency modulation $r(t)$; the intermediate case in which the modulation period is comparable to the phase slip period is of particular interest and is also investigated here in detail.

The nonautonomous Adler equation arises in a number of applications. First and foremost it arises in systems of driven identical active rotators \cite{ShinomotoKuramoto1986,Colet2010}, or, equivalently, driven arrays of Josephson junctions \cite{Watanabe1994constants}, described by the equations
\begin{equation}
\dot{\phi_j}=\omega(t) - \sin\phi_j  - K\sum_{m=1}^M \sin (\phi_j-\phi_m).
\label{eq:ar}
\end{equation}
Here $\omega$ is the intrinsic frequency and $K$ measures the coupling strength. In terms of the Kuramoto order parameter, $R\exp i\Phi\equiv \sum_{m=1}^M\exp i\phi_m$, this system can be written in the equivalent form
\begin{equation}
\dot{\theta_j}=\omega(t)-\dot{\alpha}-K{\tilde R}\sin\theta_j,
\label{eq:armf}
\end{equation}
where $\theta_j\equiv\phi_j-\alpha$, $K{\tilde R}=\sqrt{1+(KR)^2+2KR\cos\Phi}$ and $\tan\alpha=KR\sin\Phi(1+KR\cos\Phi)^{-1}$. Since $R$ and $\dot{\Phi}$ are in general functions of time \cite{choi1994periodic} the quantities ${\tilde R}$ and $\alpha$ will also be functions of time and these are determined by the collective dynamics of the remaining $M-1$ oscillators. When $M$ is large the latter are unaffected by the behavior of an individual oscillator, and ${\tilde R}$ and $\alpha$ can therefore be assumed to be given. The dynamics of each oscillator are thus described by an Adler equation with a time-dependent effective frequency and a time-dependent effective coupling constant. The latter dependence can be removed using the simple substitution $d\tau=K{\tilde R}\,dt$ provided $K(t)$ remains bounded away from zero. 

Nonautonomous effects also arise in phase-coupled oscillator systems of Kuramoto type \cite{kuramoto1975self} and these are of interest in neural models. In models of this type the coupling strength $K_{jk}$ between oscillators $j$ and $k$ is taken to be a function of time, reflecting either evolution of the network \cite{CuminUnsworth2007,SoBarreto2008,SoBarreto2011,PetkoskiStefanovska2012,Lee2012} or the effects of a drug, during anesthesia, for example \cite{Sheeba2008}. The simplest model of this type, 
\begin{equation}
\dot{\phi_j}=\omega - K(t)\sum_{m=1}^M \sin (\phi_j-\phi_m),
\label{eq:kuramotoK}
\end{equation}
can be written in the equivalent form
\begin{equation}
\dot{\theta}_j=\omega-\dot{\Phi}-K(t)R(t)\sin\theta_j,
\label{eq:mf}
\end{equation}
where $\theta_j\equiv\phi_j-\Phi$. Thus the dynamics of each individual oscillator are determined by the global behavior of the system through the quantities $KR$ and $\Phi$. When $M$ is large both $R$ and $\Phi$ may be taken as given, independent of the behavior of the oscillator $j$. The resulting system can be cast in the form 
\begin{equation}
\theta'_j=\tilde{\omega}(\tau) - \sin\theta_j,
\label{eq:kuramotoOmega}
\end{equation}
where the prime denotes differentiation with respect to $\tau$, $d\tau=KR\,dt$ and 
$\tilde{\omega}(\tau)=[\omega/K(\tau)R(\tau)]-\Phi'(\tau)$, with $K(\tau)\equiv K[t(\tau)]$, $R(\tau)\equiv R[t(\tau)]$ etc. It suffices, therefore, to consider the effects of a time-dependent effective frequency only. Related models arise in systems with frequency adaptation \cite{taylor2010spontaneous}. An excellent review of the origin of nonautonomous effects in the Kuramoto model and its variants can be found in \cite{clemson2013coupled}.


Finally, the nonautonomous Adler equation also describes a single resistively shunted Josephson junction driven by a biased AC current \cite{likharev1986dynamics}. Theoretical investigations of this equation, motivated by observations of Shapiro steps~\cite{shapiro1967microwave} in the supercurrent, have illuminated a wealth of mode-locking behavior~\cite{russer1972influence,renne1974analytical,abidi1979dynamics}. Large arrays of coupled Josephson junctions are thus amenable to the same type of analysis as active rotator systems~\cite{Watanabe1994constants,wiesenfeld1998frequency}.

The paper is organized as follows. In the next section we summarize the basic properties of the Adler equation with and without time-periodic modulation. In Sec.~\ref{sec:po} we study, under a variety of conditions, periodic orbits of the nonautonomous Adler equation that take the form of oscillations about a phase-locked state. In Sec.~\ref{sec:winding} we study the so-called phase-winding trajectories describing repeated phase slips and identify the regions in parameter space where different states of this type are found. In Sec.~\ref{sec:adiabatic} we show that an adiabatic analysis describes accurately the resulting parameter space not only for low modulation frequencies but in fact remains accurate far outside of this regime. Section~\ref{sec:concl} provides a brief summary of the results and discusses potential applications of the theory.

\section{The Adler equation}
\label{sec:adler}


The Adler equation (\ref{eq:adler}) with constant $r$ has several symmetries of interest. The equation is invariant under complete rotations $\mathcal{W}: \theta \rightarrow \theta + 2  \pi$, and time translations $\mathcal{T}_\tau: t \rightarrow t+\tau$ by an arbitrary real $\tau$. In addition, it is invariant under the phase symmetry $\mathcal{P}_0: (t,\theta) \rightarrow  (-t,\pi-\theta)$ and the parameter symmetry $\mathcal{R}_0: (r,\theta) \rightarrow - (r,\theta)$. As already mentioned, the fixed points or equilibria of Eq.~(\ref{eq:adler}) correspond to phase-locking between the two oscillators, and these exist in the parameter interval $|r|<1$:
\begin{equation}
\theta_{eq} = \sin^{-1} r.
\end{equation}
If $\theta$ is defined mod $2\pi$, this condition determines two branches of equilibria that merge in a saddle-node bifurcation at $r=\pm 1$ and are related by $\mathcal{P}_0$. One of these branches is stable and can be identified by the condition $\partial_{r} \theta_{eq} > 0$ while the other is unstable and is characterized by $\partial_{r} \theta_{eq} < 0$. No fixed points exist for $|r|>1$: $\theta$ increases monotonically when $r>1$ and decreases monotonically when $r<-1$.  When $\theta$ is defined mod $2\pi$ the resulting trajectories are both periodic in time and the steady state SNIPER bifurcations at $r=\pm 1$ generate periodic orbits, a consequence of the global organization of the stable and unstable manifolds of the fixed points.

In the present work we find it convenient to think of $\theta$ as a variable defined on the real line. When this is done the equation has an infinite number of stable and unstable equilibria that differ in the number of $2\pi$ turns relative to an arbitrary origin $\theta=0$. We refer to these turns as phase slips since one of the two oscillators is now ahead of the other by an integer number of $2\pi$ rotations. Trajectories outside of the phase-locked region will incur positive or negative phase slips with frequency 
\begin{equation}
\omega_0=\sqrt{r^2-1}. 
\label{eq:w0}
\end{equation}
This frequency approaches zero in a characteristic square root manner as $|r|$ approaches $|r|=1$ from above \cite{strogatz2014nonlinear}.

When the frequency parameter $r$ oscillates in time,
\begin{equation}
\label{eq:tforcing}
r=r_0+a\sin(2\pi t/T),
\end{equation}
the system retains the winding symmetry $\mathcal{W}$, while the translation symmetry becomes discrete $\mathcal{T}: t \rightarrow t+ T$. The phase symmetry now includes a time shift, $\mathcal{P}: (t,\theta) \rightarrow  (T/2-t,\pi-\theta)$. The parameter symmetry takes the form $\mathcal{R}: (r_0,a,\theta) \rightarrow - (r_0,a,\theta)$. There is also an additional parameter symmetry $\mathcal{S}: (t,a) \rightarrow (t+T/2,-a)$. We remark that, as already explained, any time-dependence in the coupling parameter $K>0$ can be removed by a simple transformation, and this parameter is therefore scaled to unity.

Depending on the amplitude $a$ and the period $T$ of the frequency modulation (\ref{eq:tforcing}) the solutions of the resulting equation take the form of oscillations about a phase-locked state or describe repeated phase slips in which the phase difference $\theta$ drifts with a nonzero mean speed. We identify below a series of resonances between the modulation period $T$ and the time scale for the generation of a phase slip. The resulting parameter space structure is determined using a combination of numerical simulations, numerical continuation~\cite{doedel1981auto} and asymptotic methods. Regions with an integer number of phase slips per period are separated by regions with noninteger numbers of phase slips, and include canard trajectories that drift along unstable equilibria. Both high and low frequency modulation is considered. We do not consider noise-triggered phase slips.

\section{Periodic orbits}
\label{sec:po}

\begin{figure}
{}\hspace{1cm}(a)\hspace{7cm}(b)\\
\includegraphics[width=75mm]{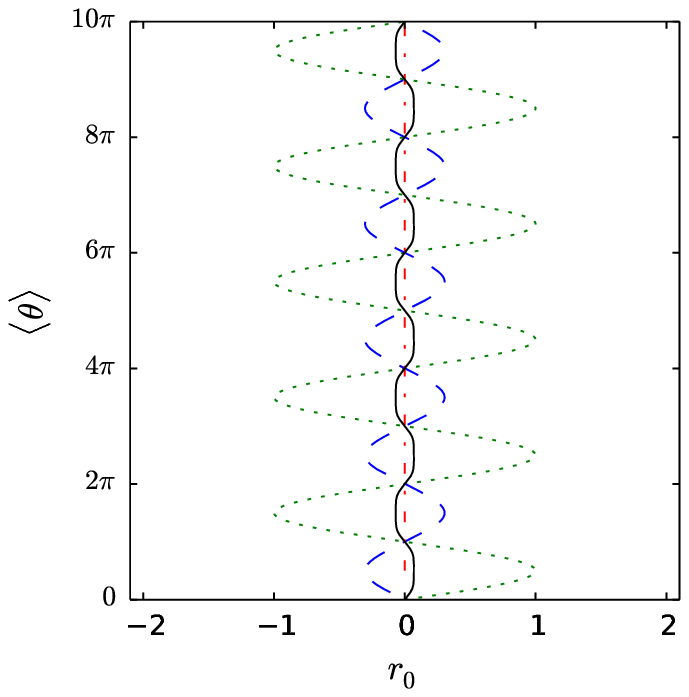}
\includegraphics[width=75mm]{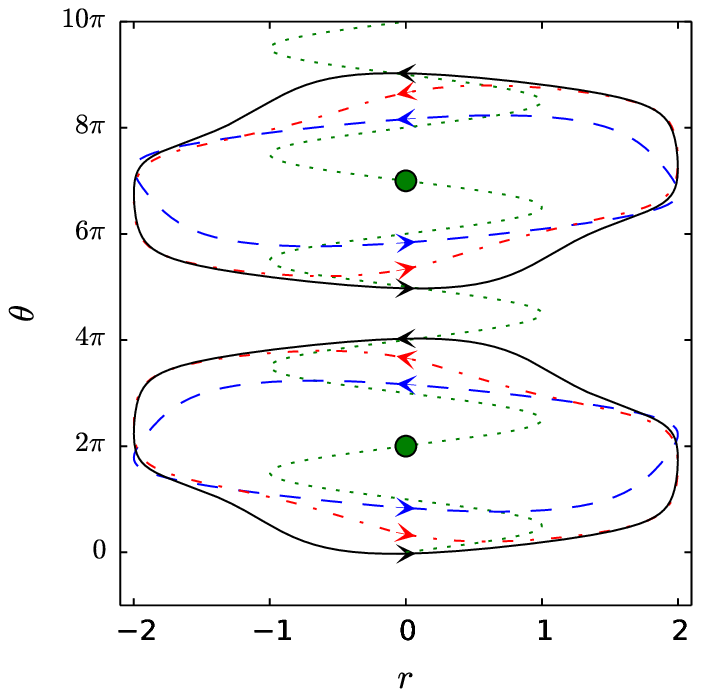}
\caption{(Color online) (a) Bifurcation diagram showing the average phase $\langle\theta\rangle\equiv T^{-1}\int_0^T\theta(t)\,dt$ of periodic orbits as a function of $r_0$ when $a=2$ and $T=15$ (blue dashed line), $T\approx 23.01$ (red dash-dotted line) and $T=25$ (black solid line). (b) Sample trajectories, in corresponding line type, in the $(r,\theta)$ plane for solutions with $r_0=0$ and $\langle\theta\rangle=2\pi$ and $7\pi$, superposed on the branch of equilibria of the autonomous system ($a=0$), represented by a green dotted line.   }
\label{fig:PObifr}
\end{figure}

Phase-locked states of the autonomous system (\ref{eq:adler}) may undergo phase slips in the presence of modulated frequency while remaining phase-locked on average. For such solutions the number of negative phase slips balances the number of positive phase slips over one modulation cycle. Figure~\ref{fig:PObifr} shows the bifurcation diagram for the nonautonomous Adler equation (\ref{eq:adler}) with the periodic modulation (\ref{eq:tforcing}) along with sample trajectories at two points on the solution branches, both superposed on the corresponding equilibrium solutions of the autonomous system, i.e., $r=r_0$. The solution branches {\it snake}, i.e., they undergo repeated back-and-forth oscillations as the parameter $r_0$ varies. The extrema of these oscillations correspond to the SNIPER bifurcations at $r=\pm1$; the equilibria with a positive slope correspond to stable solutions while those with a negative slope are unstable. Thus along the branch of equilibria stability changes at every fold.

The trajectories shown in Fig.~\ref{fig:PObifr}(b) are periodic, with period $T$, and their bifurcation structure parallels that of the phase-locked states in the autonomous system: the solutions snake within an $r_0$ interval determined by a pair of folds on either side as shown in Fig.~\ref{fig:PObifr}(a). The amplitude of this oscillation and its shape depends on the period $T$ of the forcing which also affects the solution stability. For example, for $\langle \theta\rangle=2\pi$ and $r_0 = 0$, the solution of the autonomous problem is stable, but becomes unstable for $T=15$ as most of the periodic orbit tracks the unstable branch of the autonomous problem, before becoming stable again for $T=25$. A numerical computation of the Floquet multiplier $\exp\left(-\int_0^T\cos\theta(t)dt\right)$ for the Adler equation linearized about the periodic orbit during the continuation procedure confirms that the upward (downward) sloping portions of the solution branch remain stable (unstable) all the way to the folds. 

\begin{figure}
\includegraphics[width=75mm]{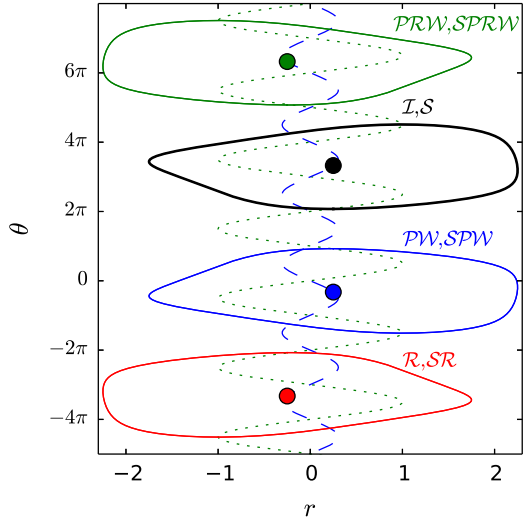}
\caption{(Color online) The four distinct orbits generated on applying the symmetries $(\mathcal{I,R,P,S})$ to the stable periodic orbit computed for $T=15$, $r_0=0.2$, and $a=2$.  A sequence of orbits with $\theta\to \theta+2\pi n$ can be found by applying $\mathcal{W}^n$ to each of the four solutions. These orbits lie on the branch displayed in Fig. \ref{fig:PObifr}(a) for $T=15$. The symmetry $\mathcal{W}$ has been applied in order to prevent overlap between the four distinct orbits. The equilibria of the autonomous system ($a=0$) are shown as a green dotted line.}
\label{fig:posym}
\end{figure}

The presence of the symmetries allows to generate other solutions from the one calculated. Figure~\ref{fig:posym} shows the four different orbits produced by applying the eight different symmetries generated by $(\mathcal{I,R,P,S})$: $\mathcal{I,R,P,S,RP,RS,PS,RPS}$ to a periodic orbit obtained for $r_0 = 0.2$, $T=15$ and $a=2$. These periodic orbits lie on the same solution branch in Fig.~\ref{fig:PObifr}(a).  The symmetry $\mathcal{S}$ acts like the identity, the time shift compensating for the reversal of $a$. Application of $\mathcal{T}$ does not produce new orbits, and we can shift any periodic orbit to higher or lower values of $\theta$ by multiples of $2\pi $ using powers of $\mathcal{W}$.  We take advantage of the latter to avoid overlap among the different solutions shown in Fig.~\ref{fig:posym}.

\begin{figure}
{}\hspace{1cm}(a)\hspace{7cm}(b)\\
\includegraphics[width=75mm]{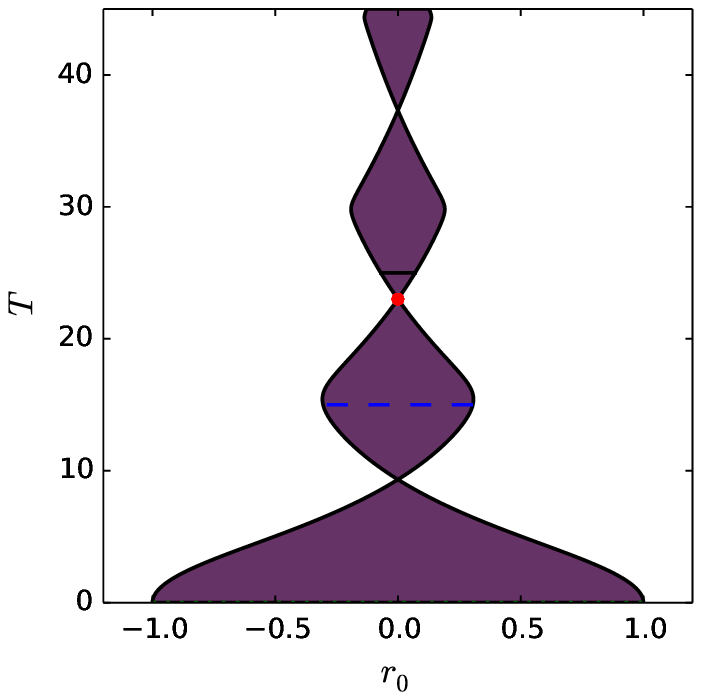}
\includegraphics[width=75mm]{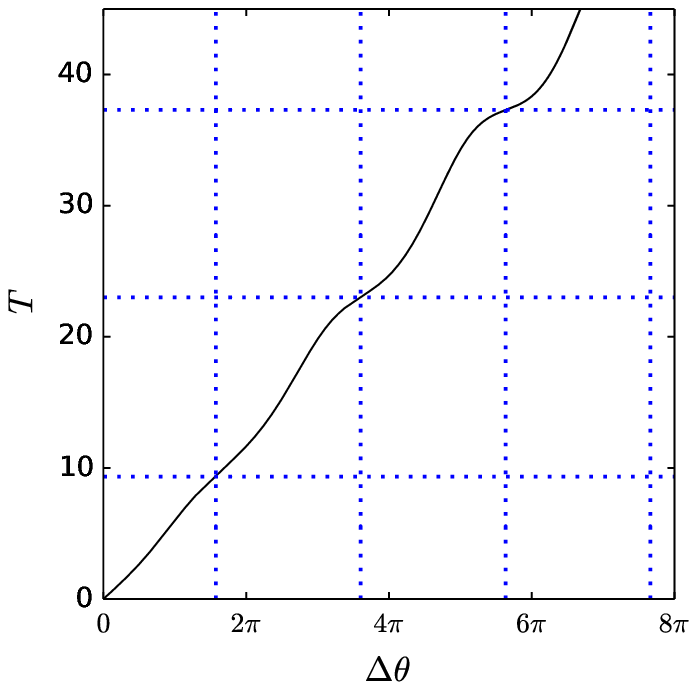}
\caption{(a) Locus of the folds that define the boundary of the $PO$ region in the $(r_0,T)$ plane. The horizontal dashed and solid lines indicate the values of $T$ corresponding to the branches of periodic orbits computed in Fig.~\ref{fig:PObifr}(a). (b) The amplitude $\Delta\theta\equiv\theta_\mathrm{max}-\theta_\mathrm{min}$ of a periodic orbit with $r_0=0$ and $a=2$ as function of the period $T$. The dotted horizontal lines correspond to the pinched zones at $T\approx 9.33$, $23.01$ and $37.31$ in panel (a); at these the corresponding periodic orbits are characterized by $\Delta \theta \approx 4.95$, $11.32$ and $17.71$ and deviate from multiples of $2\pi$ by $(2\pi n - \Delta \theta)/2\pi \approx 0.21$, $0.20$, $0.18$, respectively. } 
\label{fig:PObifrT}
\end{figure}

\begin{figure}
\includegraphics[width=75mm]{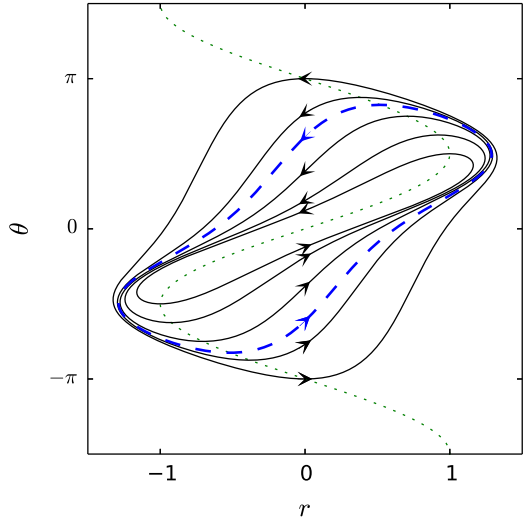}
\caption{(Color online) A series of periodic orbits (solid black) for $T=25$, $r_0=0$ and increasing values of $a$, corresponding to increasing oscillation amplitude $\Delta\theta=\pi, 5\pi/4, 3 \pi/2, 7\pi/4, 2\pi$, superposed on top of the bifurcation diagram of the phase-locked solutions of the autonomous system $a=0$ (green dotted line). The transition from zero phase slips to one positive and one negative phase slip is indicated by a dashed blue line and corresponds to $a\approx 1.29$ and $\Delta \theta \approx 1.65\pi$. 
} 
\label{fig:phaseslip}
\end{figure}

Figure~\ref{fig:PObifrT} shows how the existence region of the periodic orbit, labeled $PO$, evolves with $T$. 
Numerical continuation of the folds at the two edges of $PO$ reveals a series of {\it pinched zones} in which the folds ``cross'' and the $PO$ region is reduced to the single value $r_0=0$. This accounts for the switch in the orientation of the branch as $T$ increases (see Fig.~\ref{fig:PObifr}(a)). We call the regions between the pinched zones {\it sweet spots}. Within each of these sweet spots, the number of positive and negative phase slips during one cycle is the same, and the orbits are therefore qualitatively similar.  The resulting structure, shown in Fig.~\ref{fig:PObifrT}(a), is reminiscent of the structure observed in \cite{gandhi2015localized}. Figure~\ref{fig:PObifrT}(b) shows the amplitude of the oscillation in $\theta$ for periodic orbits at $r_0 = 0$ as a function of the period $T$. The figure reveals that $N$ positive and negative phase slips can occur even when $\Delta \theta=\theta_\mathrm{max}-\theta_\mathrm{min}< 2\pi N$. This is a consequence of the fact that the two successive saddle-nodes at $r=\pm1 $ are separated by a phase difference of $\pi$. Figure~\ref{fig:phaseslip} shows a series of periodic orbits that transition from zero to one positive and one negative phase slip as $a$ (equivalently $T$) increases. 

\subsection{Birth of periodic orbits}

To understand the effect of a time-dependent frequency parameter on the dynamics of phase-locked oscillators, we start out by considering the high-frequency modulation limit of the Adler equation (\ref{eq:adler}) with the time-periodic modulation (\ref{eq:tforcing}). We write $T=2\pi\epsilon/ \omega$, where $\epsilon\ll 1$, and $\omega\sim \mathcal{O}(1)$ is a suitably scaled frequency, and define the fast time $\phi$ by $\omega t =\epsilon \phi$. The Adler equation becomes
\begin{equation}
\omega \partial_\phi \theta =\epsilon (r_0+a \sin\phi - \sin\theta-\partial_t\theta).
\end{equation}
We assume that $\theta(\phi,t)=\theta_0(\phi,t)+\epsilon \theta_1(\phi,t)+\epsilon^2 \theta_2(\phi,t) + \dots$ and carry out the calculation order by order. The leading order equation shows that $\theta_0=\psi_0(t)$ is independent of the fast oscillation time. The $\mathcal{O}(\epsilon)$ equation yields, after integration over the fast period of the forcing, 
\begin{equation}
\partial_t \psi_0 = r_0 - \sin\psi_0.
\end{equation}
Thus, at leading order, the averaged system follows an autonomous Adler equation with constant forcing  equal to the average of the periodically modulated case. 

The solution at order $\epsilon$ reads
\begin{equation}
\theta_1 (\phi,t) \equiv -\displaystyle\frac{a}{\omega} \cos \phi + \psi_1(t),
\end{equation}
where $\psi_1$ is determined through the solvability condition at the next order.
This next order equation reads
\begin{equation}
\omega \partial_\phi \theta_2 = -\theta_1 \cos\theta_0- \partial_t \theta_1,
\end{equation} 
and integration over the fast period gives the solvability condition
\begin{equation}
\partial_t\psi_1=-\psi_1 \cos\psi_0.
\end{equation} 

The solution at order $\epsilon^2$ is thus
\begin{equation}
\theta_2(\phi,t)=\displaystyle\tfrac{a}{\omega^2}\sin\phi\cos\psi_0(t)+\psi_2(t),
\end{equation} 
while the order $\epsilon^3$ equation reads
\begin{equation}
\omega \partial_\phi \theta_3 = -\theta_2 \cos\theta_0+\displaystyle\tfrac{1}{2}\theta_1^2\sin\theta_0- \partial_t \theta_2,
\end{equation} 
leading to a solvability condition for $\psi_2$:
 \begin{equation}
\partial_t\psi_2+\psi_2\cos\psi_0=\displaystyle\tfrac{a^2}{4\omega^2}\sin\psi_0+\displaystyle\tfrac{1}{2}\psi_1^2\sin\psi_0.
\end{equation} 

To study the average dynamics, we define the period-averaged phase 
\begin{equation}
\psi=(2\pi)^{-1}\int_0^{2\pi} \left( \theta_0+\epsilon\theta_1+\epsilon^2\theta_2 \right) d\phi.
\end{equation}
This expression is accurate to order $\mathcal{O}(\epsilon^2)$. Summing the solvability conditions now yields the equation
 \begin{equation}
\partial_t\psi=r_0-\left(1-\displaystyle\tfrac{a^2 T^2}{16\pi^2}\right)\sin\psi+\mathcal{O}(T^3),
\label{eq:addlerHF}
\end{equation}
where we have replaced $\omega / \epsilon$ by $2 \pi/ T$. Thus, in the high-frequency limit, the averaged dynamics follows an Adler equation for which the amplitude of the nonlinear term that characterizes the coupling strength between the two oscillators decreases in proportion to $(aT)^2$.  The phase-locked region of the averaged equation (\ref{eq:addlerHF}) that defines the $PO$ region for the time-dependent Adler equation thus exists for $|r_0|=1- (aT/4\pi)^2$, and the introduction of high-frequency modulation narrows the width of the phase-locked region in the parameter $r_0$ by $2(aT / 4\pi)^2$. 

\subsection{Death of periodic orbits}

Asymptotic analysis near the folds that define the edges of $PO$ can provide some insight into the break-up of the periodic orbits.  We consider perturbations about the marginally stable orbit at the left ($r_0=r_-$) and right ($r_0=r_+$) edges of $PO$ for a given modulation frequency $\omega=2\pi/T$ and amplitude $a$, namely Eq.~(\ref{eq:adler}) with $r=r_0 +a\sin \omega t$, where $r_0=r_\pm +\epsilon^2\mu$ and $\epsilon \ll 1$.  We use multiple time scales by introducing a slow time $\tau=\epsilon t$ on which the system incurs net phase slips and expand the phase variable as $\theta=\theta_0+\epsilon \theta_1 +\epsilon^2\theta_2 +\dots$.

\begin{figure}
\includegraphics[width=75mm]{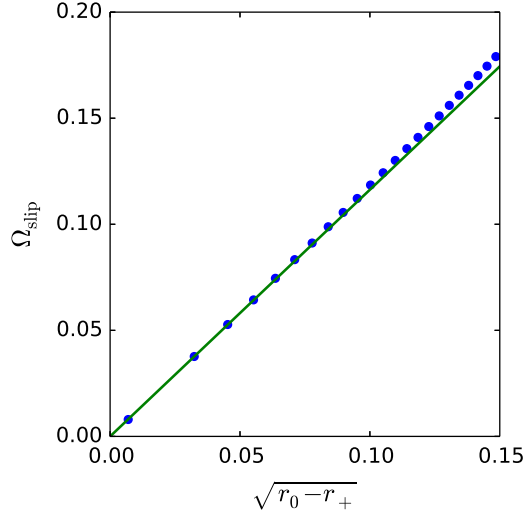}
\caption{(Color online) (a) A plot of the frequency $\Omega_\mathrm{slip}$ at which phase slips occur just outside of $PO$ as a function of the distance $\sqrt{r_0-r_+}$ from the edge when $a=2$ and $T=15$.  The solid green line is the prediction in Eq.~(\ref{eq:depinning}) from asymptotic theory while the blue dots are computed from time simulations.}
\label{fig:depinning}
\end{figure}

The leading order equation, $\partial_t\theta_0 =r_\pm + a \sin \omega t -\sin\theta_0$, is solved by the marginally stable periodic orbit, which we have computed numerically via continuation.  The $\mathcal{O}(\epsilon)$ equation is
\begin{equation}
\partial_t\theta_1+\theta_1\cos\theta_0=-\partial_\tau \theta_0
\end{equation}
which has a solution of the form $\theta_1=A \exp\left(-\int \cos\theta_0dt\right)$ for a slowly-varying amplitude $A$ as $\theta_0$ does not depend on the slow time.  At $\mathcal{O}(\epsilon^2)$, the equation reads
\begin{equation}
\partial_t\theta_2+\theta_2\cos\theta_0=\mu +\frac{1}{2}\theta_1^2\sin\theta_0-\partial_\tau \theta_1.
\end{equation}
The existence of a solution in $\theta_2$ that is $T$-periodic requires that the solvability condition 
\begin{equation}
\partial_\tau A=\mu\alpha_1 +\frac{1}{2}\alpha_2 A^2
\end{equation}
be satisfied where the coefficients can be computed numerically from the integrals
\begin{equation}
\alpha_1=\frac{1}{T}\int_0^T \exp\left(\int \cos\theta_0 dt\right)dt, \qquad \alpha_2=\frac{1}{T}\int_0^T \sin\theta_0 \exp\left(-\int \cos\theta_0 dt\right)dt.
\end{equation}
Thus, just outside of $PO$, the system will incur \textit{net} phase slips with a frequency 
\begin{equation}
\Omega_\mathrm{slip}=\sqrt{2|\alpha_1\alpha_2(r_0-r_\pm)|}.
\label{eq:depinning}
\end{equation}
Figure~\ref{fig:depinning} shows a comparison of this frequency as a function of $r_0$ with simulations near the right edge of $PO$ for $T=15$, where $r_+\approx 0.305$, and $\alpha=\sqrt{2|\alpha_1\alpha_2|}\approx 1.163$.  The coefficient that describes the square root dependence of the frequency on the distance from the left edge of $PO$ will be identical to the one computed for the right edge owing to the symmetry $\mathcal{R}$. 

\subsection{The asymptotics of sweet spots}

\begin{figure}
{}\hspace{1cm}(a)\hspace{7cm}(b)\\
\includegraphics[width=75mm]{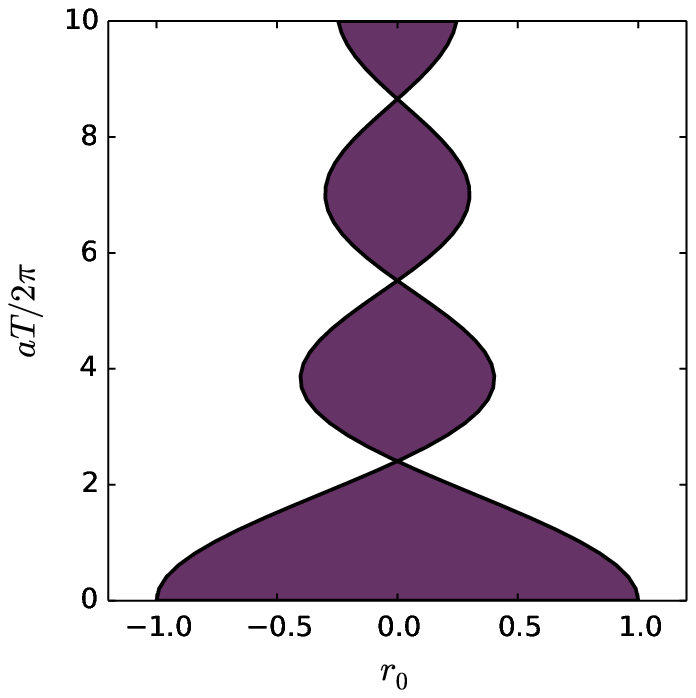}
\includegraphics[width=75mm]{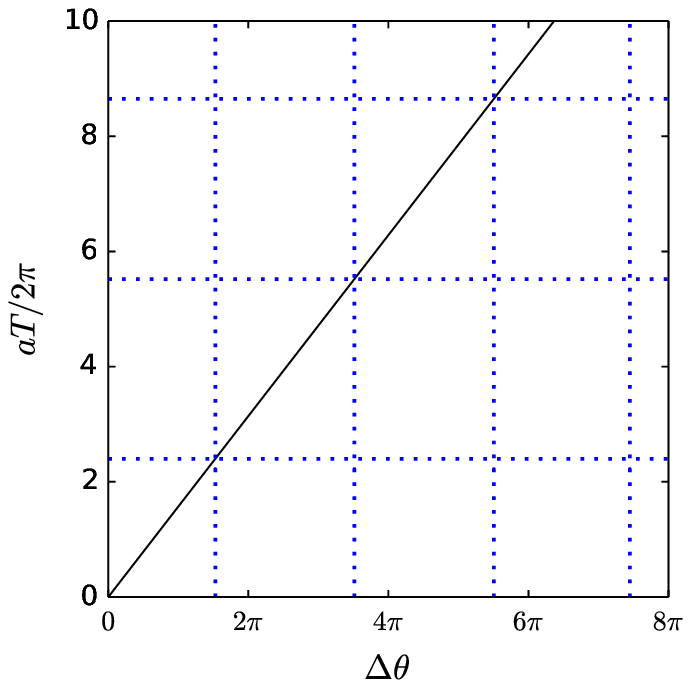}
\caption{ (a) The $PO$ region in the $(r_0,a T)$ parameter plane corresponding to stable phase-locked solutions of the Adler equation when the forcing has high frequency and a large amplitude. (b) The leading order amplitude $\Delta\theta\equiv\theta_\mathrm{max}-\theta_\mathrm{min}$  of a periodic orbit at $r_0=0$ as a function of $aT/2\pi$. Horizontal dotted lines correspond to the first three pinched zones which coincide with the zeros of $J_0$: $aT/2\pi \approx 2.40$, $5.52$ and $8.65$. } 
\label{fig:BesselHFLA}
\end{figure}

When large excursions of the forcing parameter are allowed during a high-frequency cycle, a balance is struck that allows a finite number of phase slips to occur. We keep $T=2\pi\epsilon/\omega$ but link the amplitude of the forcing to the frequency by $a=\rho/\epsilon \equiv 2 \pi \rho / \omega T$. Upon defining the fast time-scale $\phi=\omega t/\epsilon$, the Adler equation becomes
\begin{equation}
\omega \partial_\phi \theta- \rho \sin\phi =\epsilon (r_0 - \sin\theta- \partial_t\theta).
\end{equation} 
Using an asymptotic series of the form $\theta(\phi,t)=\theta_0(\phi,t)+\epsilon \theta_1(\phi,t)+\epsilon^2 \theta_2(\phi,t) + \dots$ and solving the leading order equation we obtain
\begin{equation}
\theta_0 (\phi,t)=-\tfrac{\rho}{\omega}\cos\phi + \psi_0(t).\label{eq:rho}
\end{equation}
The evolution of $\psi_0$ is determined from a solvability condition at next order. Since the order $\epsilon$ equation reads
\begin{equation}
\omega \partial_\phi \theta_1 =r_0+ \sin\left(\tfrac{\rho}{\omega}\cos\phi-\psi_0\right) - \partial_t \psi_0,
\end{equation} 
the required solvability condition is
\begin{equation}
\partial_t \psi_0= r_0-J_0(\tfrac{\rho}{\omega})\sin\psi_0,
\label{eq:adlerHFLA}
\end{equation} 
where $J_0$ is the Bessel function of the first kind. The averaged dynamics thus follow an autonomous Adler equation with a constant frequency and a coupling strength given by $J_0(\rho/\omega)=J_0(a T/2\pi)$. The boundaries of the $PO$ region are thus defined by $r_0=\pm J_0(a T/2\pi)$ and these oscillate in $r_0$ as $aT$ increases with an amplitude that decreases with increasing $T$ (Fig.~\ref{fig:BesselHFLA}).  
The location of the pinched zones is thus determined by the zeros of $J_0(aT/2\pi)$. Between these are the sweet spots where periodic orbits exist over the finite range $|r_0| < |J_0(a T/2\pi)|$. The reversal of orientation of the folds seen in Fig.~\ref{fig:PObifr}(a) is analogous to sign changes of $J_0(a T/2\pi)$ in this high frequency, large amplitude limit, as shown in Fig.~\ref{fig:BesselHFLA}.

\subsection{Amplitude dependence}

\begin{figure}
{}\hspace{1cm}(a)\hspace{7cm}(b)\\
\includegraphics[width=75mm]{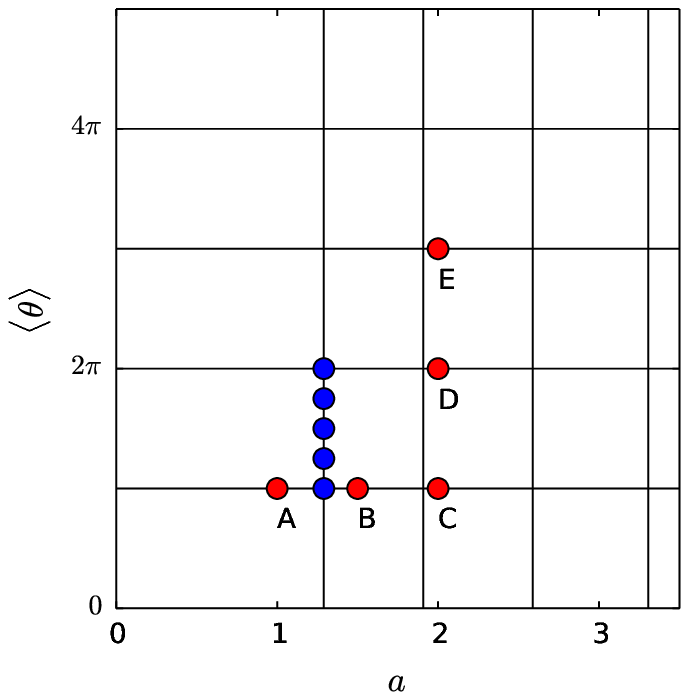}
\includegraphics[width=75mm]{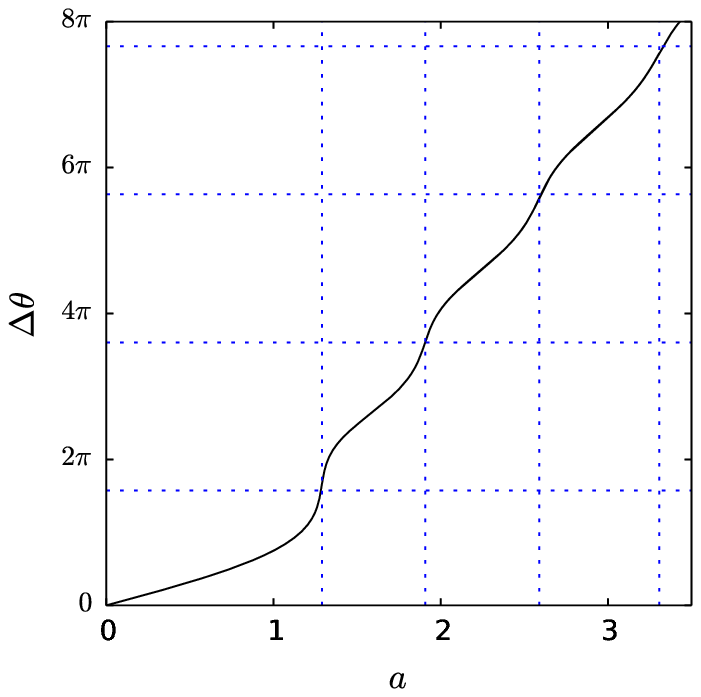}\\
\vspace{-.5cm}
{}\hspace{1cm}(c)\hspace{7cm}(d)\\
\includegraphics[width=75mm]{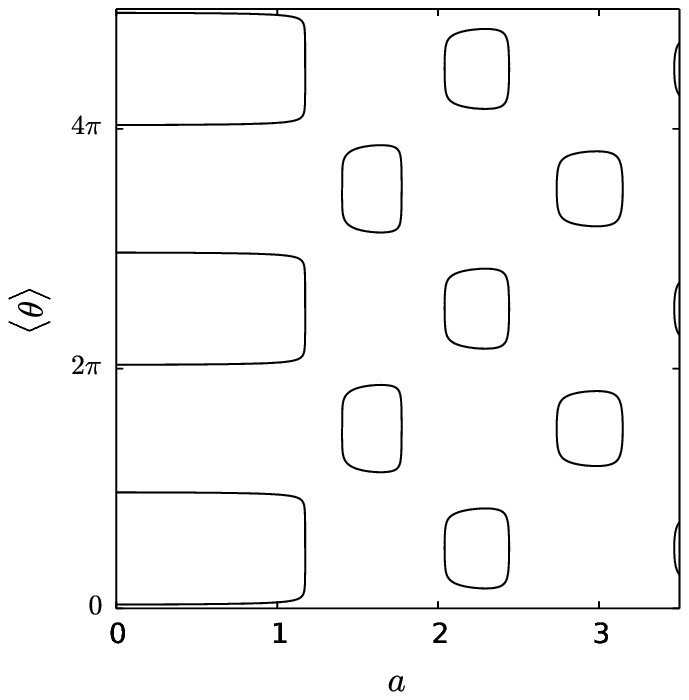}
\includegraphics[width=75mm]{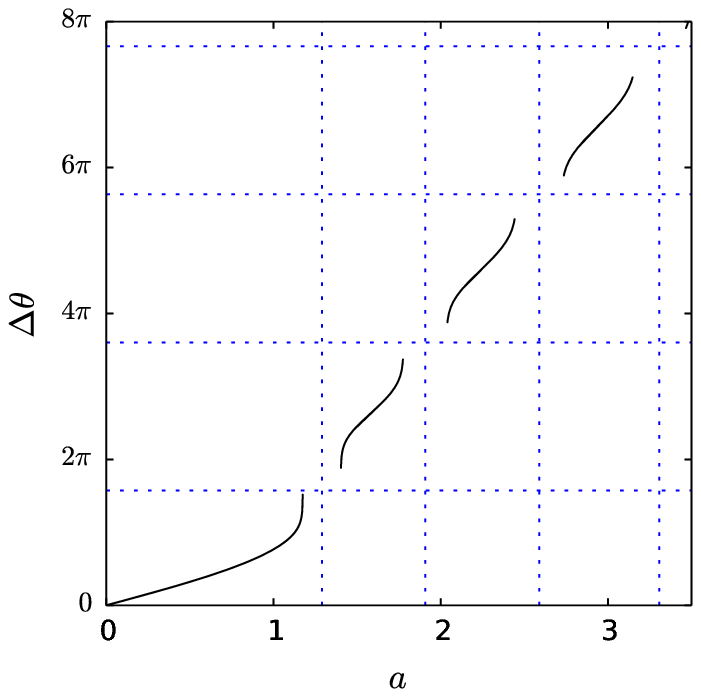}
\caption{Bifurcation diagrams showing (a,c) the average phase $\langle \theta\rangle\equiv T^{-1}\int_0^T\theta(t) dt$ (solid lines) and (b,d) the oscillation amplitude $\Delta\theta\equiv\theta_\mathrm{max}-\theta_\mathrm{min}$ of periodic orbits as a function of $a$ when $r_0=0$ and $T=25$. The solutions shown in (a) collapse onto a single curve when plotted in terms of $\Delta\theta$ in (b). When $r_0=0.1$ and $T=25$, the grid structure of  (a) separates into isolated loops shown in (b) that collapse onto disconnected line segments when plotted in terms of $\Delta\theta$ in (d).}  
\label{fig:raplanePO}
\end{figure}

We now examine how periodic solutions within $PO$ behave as a function of the amplitude of the modulation by fixing $r_0=0$ and performing numerical continuation in $a$ (Fig.~\ref{fig:raplanePO}).
As long as $r_0$ is in the interior of $PO$, each value of $a$ admits two periodic orbits on a $2\pi$ interval for $\langle \theta \rangle$. One is stable, one is unstable, and they are related by the phase symmetry $\mathcal{P}$.   The symmetries of the system further imply that the locations of these orbits at $r_0=0$ are fixed at $\langle \theta\rangle= m\pi$ for $m\in \mathbb{Z}$ and such solutions persist for all values of $a$ (horizontal lines in panel (a) of Fig.~\ref{fig:raplanePO}).
The pinched zones where the $PO$ boundaries cross (Fig.~\ref{fig:PObifrT}(a)) and the snaking branch becomes vertical (red dash-dotted line in Fig.~\ref{fig:PObifr}(a)) correspond to codimension two points in the $(a,T)$ plane; at these points a continuum of periodic orbits parametrized by the phase average $\langle \theta \rangle$ is present. Thus when $r_0=0$ the periodic orbits create the grid-like bifurcation diagram shown in Fig.~\ref{fig:raplanePO}(a). This grid structure breaks apart into isolated loops of solutions as soon as $r_0\neq 0$, and gaps between the regions of existence of periodic orbits begin to emerge (cf. Fig.~\ref{fig:PObifrT}(a)). The loops that emerge from the breakup of the rectangular grid structure at $r_0=0$ when $r_0\neq 0$ shrink to zero with increasing $a$ (or $T$), as expected from Fig.~\ref{fig:PObifrT}(a). Numerical continuation of the boundary of the $PO$ region as a function of $a$ when $r_0=0.1$ and $T=25$ reveals that periodic orbits persist only to $a\approx 14.5$.

\begin{figure}
\includegraphics[width=150mm]{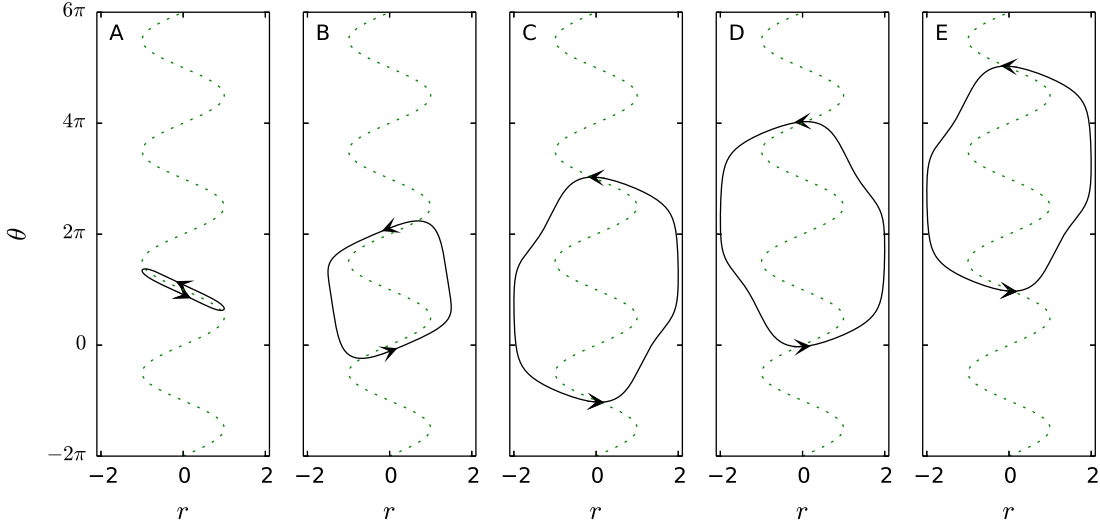}
\caption{(A)-(C) Periodic orbits with $\langle \theta\rangle = \pi$ in the $(r,\theta)$ plane when $r_0=0$, $T=30$ and $a=1,\;1.5,\;2$. (C)-(E) Periodic orbits with $\langle \theta\rangle = \pi,\;2\pi,\;3\pi$ in the $(r,\theta)$ plane when $r_0=0$, $T=30$ and $a=2$. The orbits correspond to the red dots in Fig.~\ref{fig:raplanePO}(a) labeled with capital letters.}  
\label{fig:ra_sols}
\end{figure}

Figure~\ref{fig:ra_sols} shows solutions for $r_0=0$ with parameters values indicated in Fig.~\ref{fig:raplanePO}(a) by red dots labeled with the corresponding capital letter.
The equilibria for the autonomous problem are shown for reference (dotted line). These reveal that the periodic orbits alternately track branches of unstable and stable equilibria for part of each oscillation cycle (orbits A, B, C), and likewise for C, D, E. Since orbits that track stable equilibria are expected to be stable when the drift along such equilibria is sufficiently slow, we expect that orbits B and D are stable while A, C and E are unstable. This expectation is confirmed by explicit stability calculations.

\subsection{Canards}
\label{sec:canard}

\begin{figure}
\includegraphics[width=150mm]{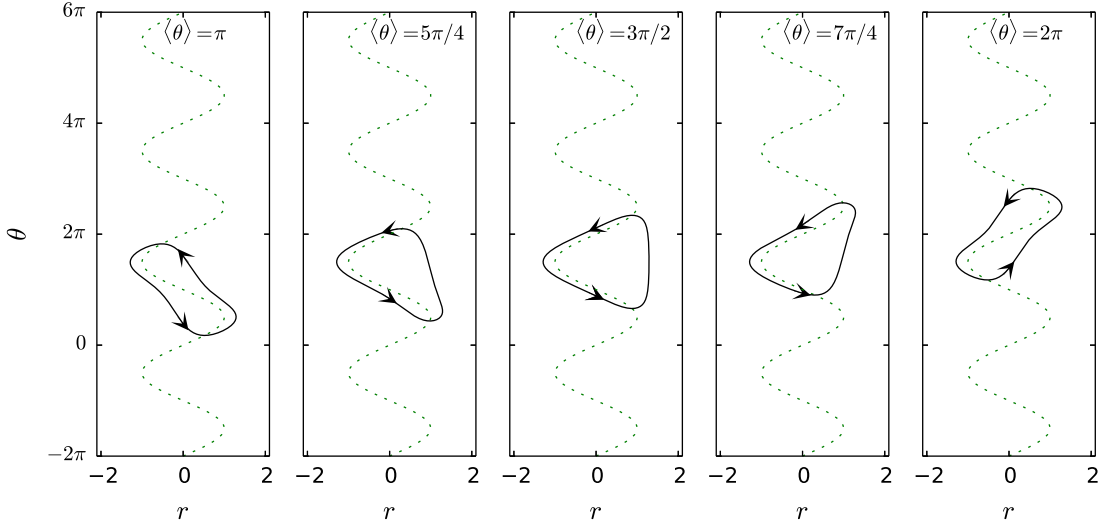}
\caption{Periodic orbits along the first vertical solution branch in Fig.~\ref{fig:raplanePO} in the $(r,\theta)$ plane when $r_0=0$, $T=25$ and $a\approx 1.2$. These solutions are characterized by $\langle\theta\rangle$ that is a fraction of $2\pi$, viz. $\pi$, $5\pi/4$, $3\pi/2$, $7\pi/4$ and $2\pi$. The orbits correspond to the unlabeled blue dots in Fig.~\ref{fig:raplanePO}(a).}  
\label{fig:ra_solsC}
\end{figure}

\begin{figure}
{}\hspace{1cm}(a)\hspace{7cm}(b)\\
\begin{minipage}{0.48\columnwidth}
\includegraphics[width=75mm]{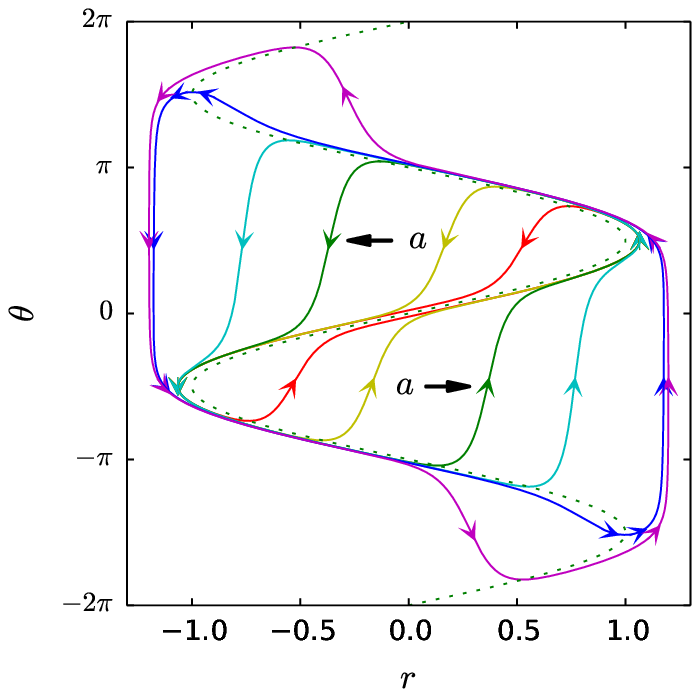}
\end{minipage}
\begin{minipage}{0.48\columnwidth}
\includegraphics[width=75mm]{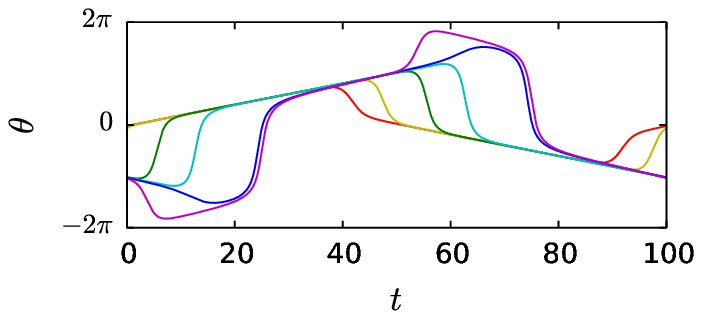}\\
\includegraphics[width=75mm]{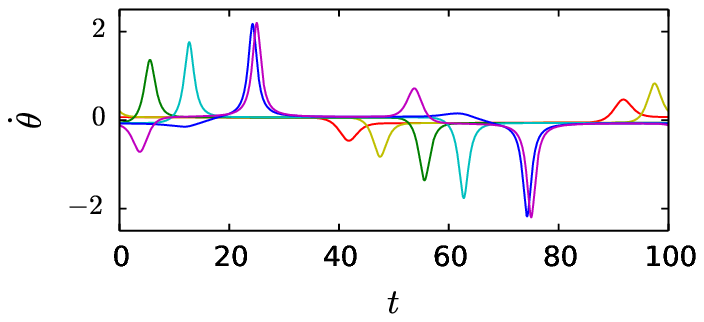}
\end{minipage}
\caption{(Color online) (a) Two-headed canard trajectories $\theta(r)$ for $r_0=0$, $T=100$ and $a\approx 1.064807$, 1.064865, 1.064871, 1.064872, 1.064876, 1.066086, 1.177531 and $1.198182$. 
 (b) The corresponding solutions $\theta(t)$ and $\dot{\theta}(t)$.}
\label{fig:Canard100}
\end{figure}

\begin{figure}
{}\hspace{1cm}(a)\hspace{7cm}(b)\\
\begin{minipage}{0.48\columnwidth}
\includegraphics[width=75mm]{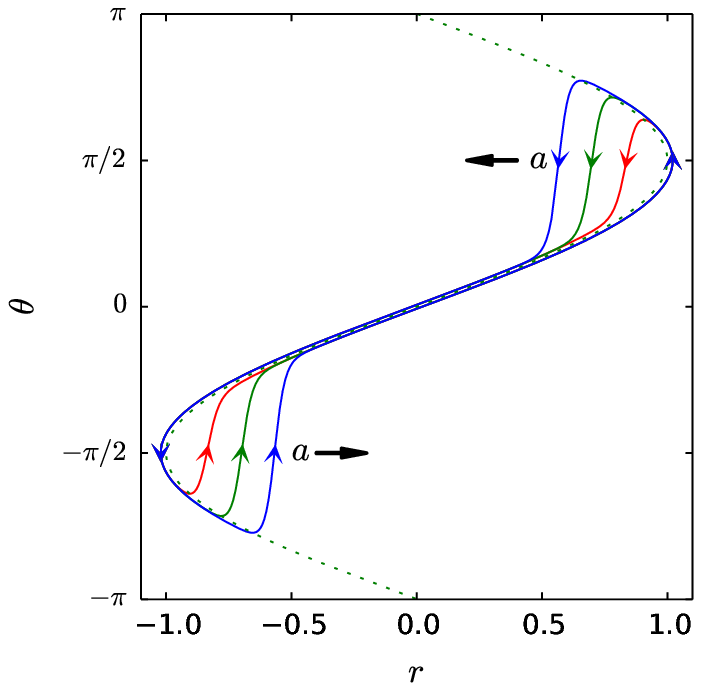}
\end{minipage}
\begin{minipage}{0.48\columnwidth}
\includegraphics[width=75mm]{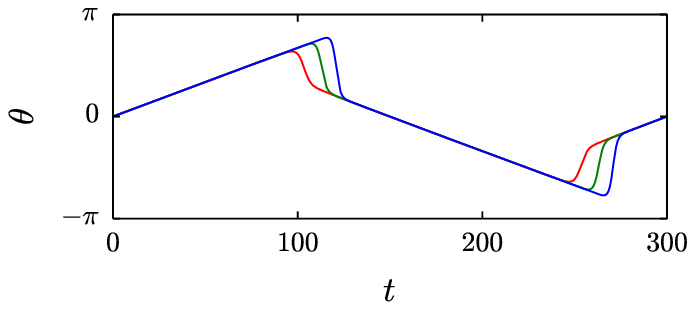}\\
\includegraphics[width=75mm]{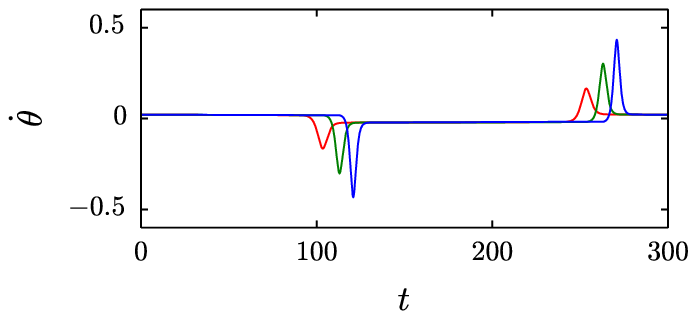}
\end{minipage}
\caption{(Color online) Two-headed canard trajectories computed by numerical continuation of periodic orbits in the parameter $a$. The parameters are $r_0=0$, $T=300$ and $a\approx 1.02115308$, 1.02116560, 1.02116562.  }
\label{fig:CanardAuto}
\end{figure}

Figure~\ref{fig:ra_solsC} shows periodic orbits from the first vertical solution branch in Fig.~\ref{fig:raplanePO}(a) corresponding to the dark blue dots not labeled with capital letters.
These periodic orbits all have the same value of $a\approx 1.2$ and correspond to pinched zone solutions with $\langle \theta\rangle= \pi$, $5\pi/4$, $3\pi/2$, $7\pi/4$ and $2\pi$. These solutions illustrate how the periodic orbit expands to larger $\langle \theta\rangle$ while tracking the equilibria of the autonomous system. These are beginning to reveal characteristics of the so-called canard states familiar from studies of slow-fast systems. For example, the third panel shows an orbit that slowly tracks a branch of stable equilibria towards lower $\theta$ and smaller $r$ followed by tracking a branch of unstable equilibria towards yet smaller $\theta$ but increasing $r$, before an abrupt transition near the right fold that restores the original $\theta$ value essentially instantaneously, i.e., at essentially constant $r$. This difference in timescales is not as pronounced in the last panel of Fig.~\ref{fig:ra_solsC} but can be enhanced by increasing the modulation period. Figure~\ref{fig:Canard100} shows typical canard trajectories with a clear separation of timescales, obtained for $T=100$, $r_0 = 0$ and slightly different modulation amplitudes $a$.
Increasing the amplitude $a$ very slightly leads the canard to overshoot the right saddle-node and can make it depart from the branch of unstable equilibria upwards, i.e., in the opposite direction as compared to the solutions for slightly smaller $a$. The latter case leads to a different type of canard: the system jumps from the unstable solution branch to the upper branch of stable equilibria, which it then follows downward in $\theta$. After reaching the upper left fold of the equilibria the trajectory jumps to the lower left fold and thereafter follows the lower unstable equilibria towards larger $r$, resulting in the same sequence of transitions but now as $r$ increases. The resulting solution is periodic but is characterized by phase slips that take place inside the snaking region $|r|<1$. This behavior is exemplified by the outer canard trajectory in Figs.~\ref{fig:Canard100}(a); the associated $\dot\theta$ displays an inverse peak, as shown in Fig.~\ref{fig:Canard100}(b). Additional time separation along the stable and unstable manifolds can be achieved by increasing $T$ further. For example, Fig.~\ref{fig:CanardAuto} shows several ``two-headed'' canard trajectories obtained for $T=300$.

\section{Winding trajectories}
\label{sec:winding}

Outside of the phase-locked region of the Adler equation with constant frequency parameter ($r_0 \in [-1,1]$, $a=0$) there exist winding solutions that complete phase slips with the frequency given by Eq.~(\ref{eq:w0}). The introduction of a modulation in $r$ with period $T$ (Eq.~\ref{eq:tforcing}, $a\ne0$) generates winding solutions even when the average value $r_0$ lies within $[-1,1]$.  This occurs for values of $r_0$ outside of $PO$ (but $|r_0|<1$), and is a consequence of an imbalance between positive and negative phase slips.

\begin{figure}
{}\hspace{1cm}(a)\hspace{7cm}(b)\\
\includegraphics[width=80mm]{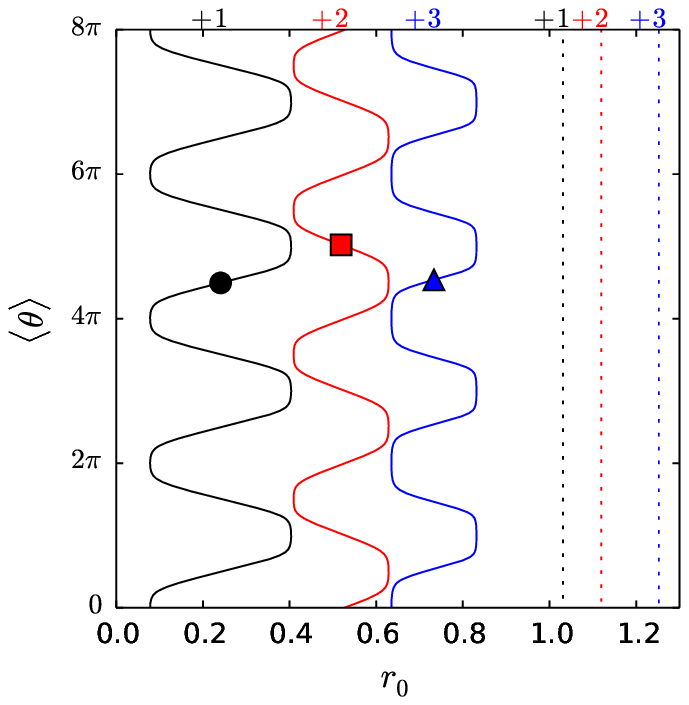}
\includegraphics[width=80mm]{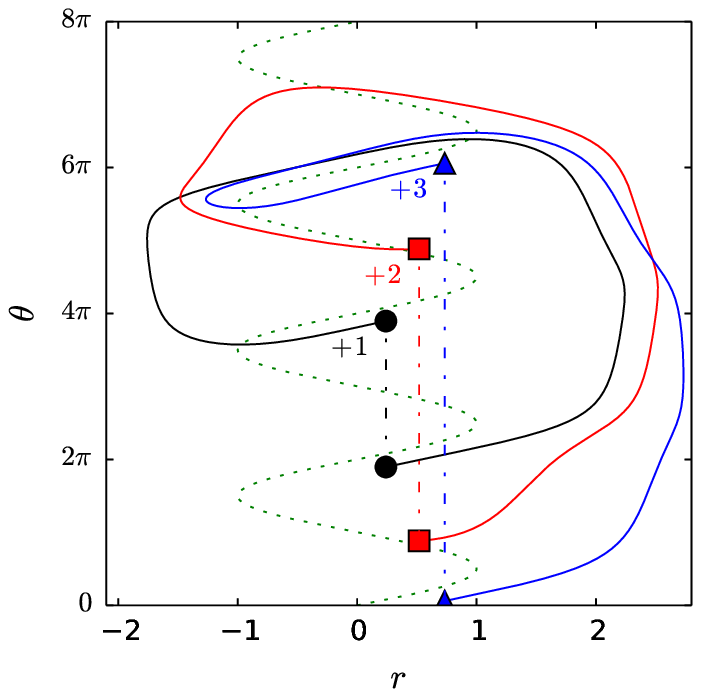}
\caption{(Color online) (a) The phase $\langle\theta\rangle\equiv T^{-1}\int_0^T\theta(t)\,dt$  averaged over $T$ of winding orbits as a function of $r_0$ when $a=2$ and $T=25$. Since $\theta$ is no longer periodic all points with the same $\langle\theta\rangle$, mod $2\pi$, at a particular value of $r_0$ lie on the same trajectory.  The black (with circle), red (with square) and blue (with triangle) branches have winding numbers $N=1,\; 2,\;3$, respectively. The branches of solutions with the same winding numbers but constant frequency parameter $r=r_0$ are shown as (vertical) dotted lines.  (b) Sample winding trajectories corresponding to the colored symbols in panel (a). }
\label{fig:sno123}
\end{figure}

We define the \textit{winding number} of a trajectory in the modulated system as the average number of net phase slips per period,  
\begin{equation}
N=\lim_{m\to \infty} \frac{\theta(mT)-\theta(0)}{2\pi m}
\end{equation}
with $m\in \mathbb{Z}$. Figure~\ref{fig:sno123} shows solution branches with integer winding numbers $N=1,2,3$ when $a=2$ and $T=25$ (solid lines).  These were computed by numerical continuation as a boundary value problem with the constraint that $\theta(T)-\theta(0)=2\pi N$.
Trajectories with integer winding number exist over finite ranges of the parameter $r_0$. Solutions displaying an extra positive phase slip over each modulation cycle have winding number $N=1$; these exist for $r_{1,{\rm min}} \approx 0.1 < r_0 < r_{1,{\rm max}} \approx 0.4$. To the right of this interval lie solutions with winding number $N=2$, extending from $r_{2,{\rm min}} \approx 0.4$ to $r_{2,{\rm max}} \approx 0.6$. Solutions with higher integer winding number exist beyond this point as exemplified by the $N=3$ solutions in Fig.~\ref{fig:sno123}.

\subsection{Resonance tongues}

The parameter range containing integer winding solutions forms through the opening of resonance tongues as the modulation amplitude $a$ increases from zero.   We write, following \cite{renne1974analytical}, $x=\tan\theta/2$ to put the Adler equation in the form
\begin{equation}
\dot{x}=\frac{1}{2}r-x+\frac{1}{2}r x^2.
\end{equation}
The Riccati transformation $x=-2\dot{y}/ry$ now generates a second order linear equation for the variable $y(t)$:
\begin{equation}
\ddot{y}+\left(1-\frac{\dot{r}}{r}\right)\dot{y}+\frac{r^2}{4}y=0.
\end{equation}
Using the standard transformation $y=ze^{-\tfrac{1}{2}\int 1-\frac{\dot{r}}{r}dt}$ we finally obtain the Hill equation
\begin{equation}
\ddot{z}+\left[\frac{r^2}{4}+\frac{\ddot{r}}{2r}-\frac{\dot{r}^2}{2r^2}-\frac{1}{4}\left(1-\frac{\dot{r}}{r}\right)^2   \right]z=0.\label{eq:hill}
\end{equation} 
Substituting the time-dependent frequency parameter $r$ specified in Eq.~(\ref{eq:tforcing}) and assuming $a\ll1$ yields the Mathieu equation
\begin{equation}
\ddot{z}+\left(\frac{r_0^2-1}{4}+\frac{a}{2 r_0} \sqrt{\omega^2+(r_0^2-\omega^2)^2}\sin(\omega t-\xi)\right)z+\mathcal{O}(a^2) =0,
\label{eq:hillmat}
\end{equation}
where $\omega\equiv 2\pi/T$ and $\tan\xi\equiv(r_0^2-\omega^2)/\omega$.  Phase slips in the original Adler equation correspond to divergences of $x$; these in turn correspond to zero crossings of $y$ and $z$. 

\begin{figure}
{}\hspace{1cm}(a)\hspace{7cm}(b)\\
\includegraphics[width=75mm]{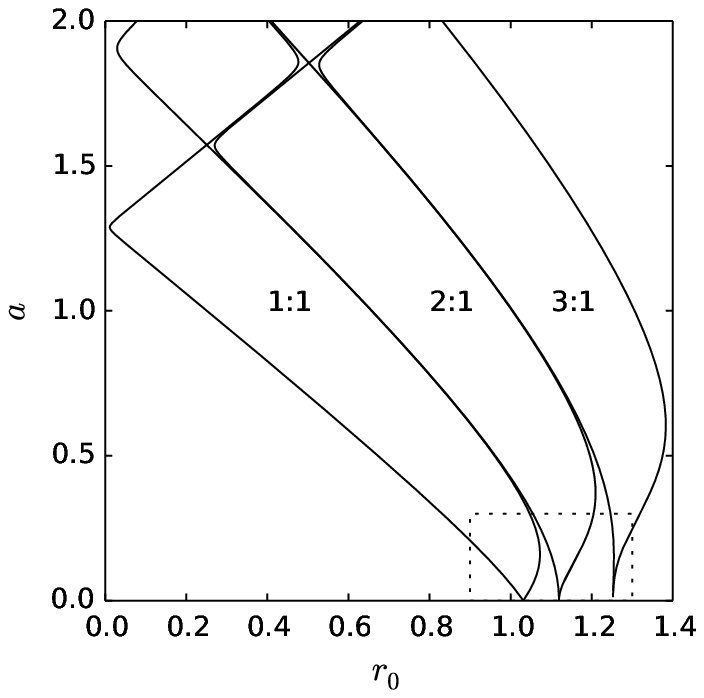}
\includegraphics[width=75mm]{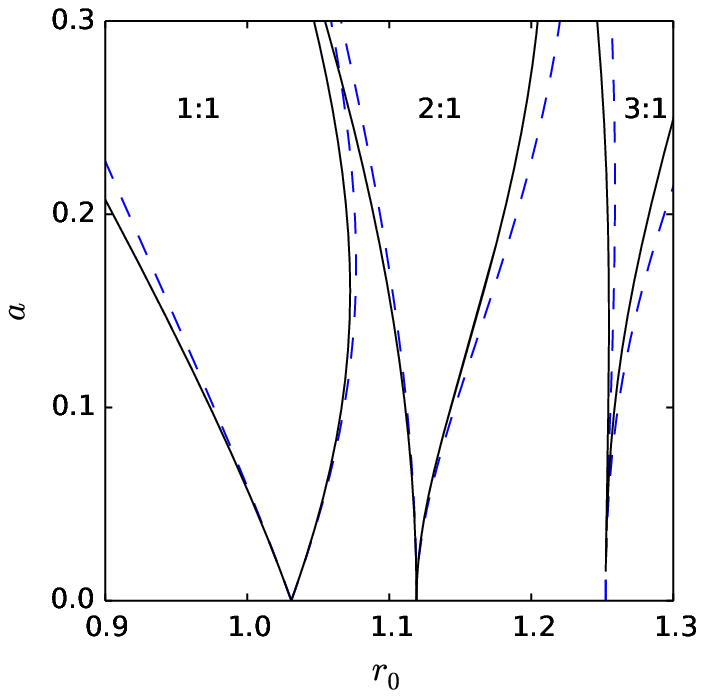}
\caption{(a) Resonance tongues for the 1:1, 2:1 and 3:1 resonances between the winding frequency and the modulation frequency in the $(r_0,a)$ plane when $T=25$. The resonance tongues correspond to the solution branches shown in Fig.~\ref{fig:sno123} with 1, 2 and 3 phase slips per period of the modulation cycle, respectively.  The boxed region in the lower right of panel (a) is replotted in panel (b) along with the predictions for the location of the tongues from Eq.~(\ref{eq:hillmat}) in dashed lines.}
\label{fig:atongue}
\end{figure}

The resonance tongues grow in this asymptotic limit according to the characteristic curves of the above Mathieu equation. We compare these asymptotic predictions with the numerical computation of the resonance tongues through two-parameter continuation of folds on the branches of winding solutions. The tongues associated with the 1:1, 2:1, and 3:1 resonances between the winding frequency and the modulation frequency are shown in Fig.~\ref{fig:atongue} alongside the predictions from the characteristic curves of the Mathieu equation (\ref{eq:hillmat}).

The resonance tongues enter farther into the phase-locked region $|r_0|<1$ as $a$ increases. We observe that as $a$ increases the location of the tongues begins to depart from the Mathieu equation predictions, as noted already in the context of Josephson junction models~\cite{likharev1986dynamics} (Ch. 11). In particular, the interaction of these tongues with their negative winding counterparts leads to qualitative changes: for $a >1.29$, the width of the 1:1 resonance tongue stops growing monotonically and its left boundary turns abruptly from $r_0 \approx 0$ to larger $r_0$; at $r_0\approx 0.25$ , $a \approx 1.57$ the tongue collapses to a single point before growing again. This situation repeats as $a$ increases and the tongue therefore describes a succession of sweet spots and pinched zones. The same behavior is observed for the subsequent resonance tongues: the 2:1 resonance tongue starts to shrink at $r_0\approx 0.25$, $a \approx 1.57$ and collapses to a point at $r_0\approx 0.50$, $a\approx 1.86$, etc.

\subsection{Partitioning of the parameter space}

\begin{figure}
\includegraphics[width=75mm]{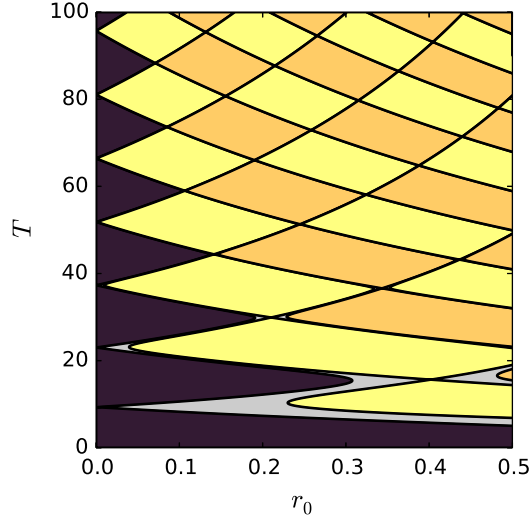}
\caption{(Color online) Average winding number per period $T$ of the frequency parameter shown in the $(r_0,T)$ plane for $a=2$. No net phase slips occur over the course of a modulation period in the dark region to the left; the alternating lighter yellow/darker orange regions to the right indicate $1,2,3,\dots$ net phase slips as $r_0$ increases. The (lightest) gray transition zones have non-integer winding numbers. Trajectories with negative winding number are located in regions obtained by reflection in $r_0=0$.}
\label{fig:rTplane}
\end{figure}

The parameter plane $(r_0,T)$ can be partitioned in terms of winding number by following the folds of the $N$:1 resonant winding trajectories such as those shown in Fig.~\ref{fig:sno123}. The resulting partitioning of parameter space is shown in Fig.~\ref{fig:rTplane}.
To obtain this figure the branches with winding numbers $1\le N\le 7$ were continued in $r_0$ for $a=2$ and $T=5$, followed by continuation of the saddle-nodes on these branches in the parameters $r_0$ and $T$. The region $PO$ of periodic orbits was computed in a similar way for completeness. The sweet spot and pinching structure of regions with constant integer winding number that begins to emerge in Fig.~\ref{fig:atongue}(a) can also be seen as $T$ increases for fixed $a$. The width of these sweet spots decreases with $T$. For infinite periods, any small departure from the symmetry axis $r_0 = 0$ leads to the dominance of positive or negative phase slips over the other.    

Thus the parameter plane is partitioned into regions with solutions displaying zero, one, two or more net phase slips per cycle. Each of these regions possesses a structure similar to that of the $PO$ region with zero net phase slips. The first region to the right of $PO$ corresponds to solutions that undergo one extra positive phase slip within each period of the modulation. The first sweet spot of this band, at low $T$, corresponds to solutions that complete one positive and no negative phase slip per cycle; the second sweet spot, further up, is comprised of solutions that complete two positive and one negative phase slips per cycle, etc. The second region on the right corresponds to solutions that undergo two extra positive phase slips, and so on as $r_0$ increases. All these regions have a similar structure as the modulation period $T$ increases. They all correspond to the resonance tongues in Fig.~\ref{fig:atongue} and are separated by transition zones with solutions that have a non-integer winding number.  These transition zones narrow as $T$ increases and solutions within them can have periods that are a multiple of the modulation period, or not be periodic at all. Solutions with negative winding number are found in analogous regions obtained by reflection in $r_0=0$.

\begin{figure}
\includegraphics[width=75mm]{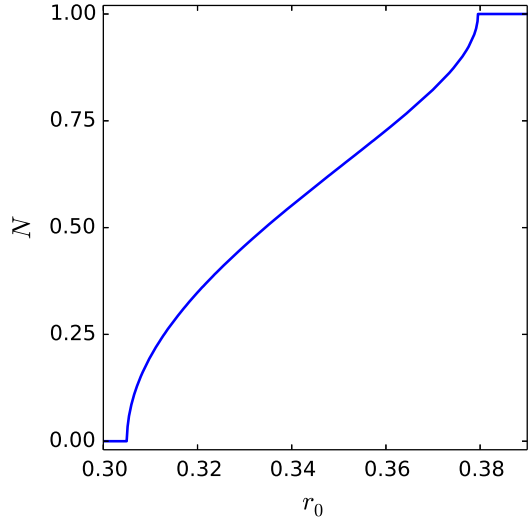}
\caption{The winding number $N$ as a function of $r_0$ across the transition zone between $PO$ and the 1:1 resonance tongue for $T=15$ and $a=2$.}
\label{fig:transition}
\end{figure}

Figure~\ref{fig:transition} shows the winding number between $PO$ and the 1:1 resonance tongue as computed from time simulations averaged over 5000 modulation periods $T=25$. The figure shows that the winding number increases monotonically and smoothly within this transition zone, as expected on the basis of theoretical considerations~\cite{renne1974analytical}. However, modifications of the nonautonomous Adler equation, such as the inclusion of an inertial term or a more general time dependence, can generate subharmonic resonances that populate the transition zones~\cite{likharev1986dynamics}. Subharmonic resonances have also been observed to produce a devil's staircase type structure in the related problem of spatially localized states in the periodically forced Swift--Hohenberg equation \cite{gandhi2014new}.

\subsection{Asymptotic formation of sweet spots}
\label{adiabB}

We can extend the above predictions by analyzing a limit in which the trajectory barely exits the phase-locking region $-1 < r < 1$ but assuming the modulation period is slow enough that phase slips still take place. Explicitly, we take $r(t)=\epsilon^2\mu+(1+\epsilon^2\rho)\sin\epsilon^2\omega t$, where $\epsilon^2\mu$ represents a small offset of the average value of $r(t)$ from $r_0=0$. We introduce the slow time scales $\tau=\epsilon t$ and $\Phi=\epsilon^2\omega t$ and employ an asymptotic expansion of the form $\theta=\theta_0+\epsilon \theta_1+\epsilon^2 \theta_2 + \dots$.

At leading order, the Adler equation (\ref{eq:adler}) gives $\sin\theta_0=\sin\Phi$ for which we choose the stable phase locked solution $\theta_0=\Phi+2\pi n$ that has no $\tau$ dependence.  The alternate choice, $\theta_0=\pi-\Phi+2\pi n$, produces unstable periodic orbits or unstable winding trajectories. At order $\epsilon$, we obtain the equation $\partial_{\tau} \theta_0=-\theta_1\cos\theta_0$. When $\theta_0\neq\pi/2+\pi n$, $\theta_1=0$ in order to satisfy the condition that $\theta_0$ be independent of $\tau$.  
At order $\epsilon^2$, we obtain
\begin{equation}
\theta_2\cos\theta_0=\mu+\rho\sin\Phi+\tfrac{1}{2}\theta_1^2\sin\theta_0-\partial_\tau\theta_1-\omega \partial_\Phi\theta_0,\label{adiab1}
\end{equation}
leading to the second order correction 
\begin{equation}
\theta_2=(\mu-\omega)\sec\Phi+\rho\tan\Phi
\end{equation}
provided that $\theta_0\neq\pi/2+\pi n$.

To examine the dynamics near $\theta_0=\pi/2 + n \pi$ where the system is transitioning between phase-locked dynamics and winding, we take the slow time to be $\Phi=\pi/2+\epsilon\phi$.  Equation (\ref{eq:adler}) then becomes
\begin{equation}
\epsilon\omega \partial_\phi \theta=\epsilon^2\mu+(1+\epsilon^2\rho)\cos\epsilon\phi - \sin\theta.
\end{equation}
The leading order and order $\epsilon$ equations are identical to the general case above while $\theta_1$ is determined from the order $\epsilon^2$ equation,
\begin{equation}
\theta_2\cos\theta_0=\mu +\rho-\tfrac{1}{2}\phi^2 +\tfrac{1}{2}\theta_1^2\sin\theta_0 - \omega \partial_\phi \theta_1,
\end{equation}
which differs from Eq.~(\ref{adiab1}). Since $\theta_0=\pi/2$ the Riccati transformation $\theta_1=-2\omega \partial_\phi \psi/\psi$ transforms this equation into the Weber equation 
\begin{equation}
\label{eq:weber}
\partial_\phi^2\psi=-\tfrac{1}{2\omega^2}(\mu+\rho-\tfrac{1}{2}\phi^2)\psi.
\end{equation}

We now use a matching procedure to connect the relevant solution of this equation to the case when $\Phi\neq \pi/2 + n \pi$. Noting that $\theta_1(\Phi)\to 0$ as $\Phi\to \pi/2 + n \pi$, we choose our solution such that $\theta_1 \to 0$ as $\phi\to -\infty$.  This matching condition is satisfied by the parabolic cylinder function $\psi=D_\nu(s)$, where
\begin{equation}
\nu=\frac{\mu+\rho}{2 \omega}-\frac{1}{2},\qquad  s=\frac{\phi}{\sqrt{\omega}}. 
\end{equation}
Each zero $s=s_0$ of $\psi=D_{\nu}(s)$ corresponds to one phase slip. 
Care must be taken in interpreting these results since the zeros of $\psi$ correspond to divergences of $\theta_1$ and thus a breakdown of the asymptotic series used to obtain Eq.~(\ref{eq:weber}). The above calculation holds between the asymptotic breakdowns where $\theta_1(\tau)$ diverges, so a complete trajectory can be constructed by ``gluing'' solutions across each individual phase slip.  Thus $\psi$ can be used to describe a series of phase slips via this gluing process. 

The number of zeros of $\psi$ corresponds to the number of phase slips and thus determines which solution branch the system will follow upon re-entering the phase-locked region $\pi/2<\Phi<3\pi/2$. In particular, $[n_+]$ phase slips are undergone when 
\begin{equation}
[n_+]-\tfrac{1}{2}<\frac{\mu+\rho}{2\omega}<[n_+]+\tfrac{1}{2}.
\end{equation}

\begin{figure}
\includegraphics[width=75mm]{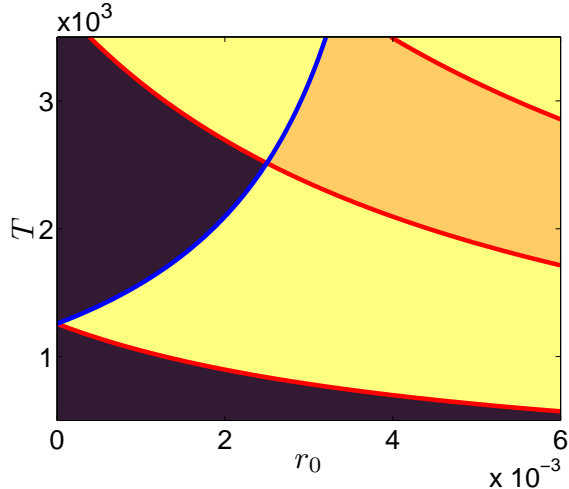}
\caption{(Color online) Average winding number per period $T$ of the frequency parameter shown in the $(r_0,T)$ plane for $a=1.005$.  Colors represent results from numerical simulation: no net phase slips occur over the course of a modulation period in the dark region ($PO$) to the left; the alternating yellow/orange regions to the right indicate $1,2,3,\dots$ net phase slips as $r_0$ increases. The red (negative slope) and  blue (positive slope) lines mark the transitions between the regions of constant $n_+$ and $n_-$ as predicted by the asymptotic theory (Eq.~(\ref{eq:asymptoticN})). Trajectories with negative winding number are located in regions obtained by reflection in $r_0=0$.}
\label{fig:asymp}
\end{figure}

More generally, we can express the number of positive (negative) phase slips that occur near the boundaries of the phase-locked region in terms of the parameters of the problem as
\begin{equation}
[n_\pm]=\begin{cases}
\pm\left[\frac{(a\pm r_0-1)T}{4\pi}\right]& (a\pm r_0-1)\geq 0\\
0 & (a\pm r_0-1)<0,
\end{cases} 
\label{eq:asymptoticN}
\end{equation} 
where the square bracket indicates rounding to the nearest integer. The predictions of this theory match well with time simulations for $a=1.005$, as seen in Fig.~\ref{fig:asymp}. The simulations employed a fourth order Runge--Kutta scheme for 12 periods of the modulation using the initial condition $\theta(0)=\sin^{-1}r_0$. The winding number was computed from $2\pi N=\left( \theta(12T)-\theta(2T)\right)/10$ as a function of the parameters $r_0$ and $T$.  Time simulations were used in place of numerical continuation because the extremely long time scales make continuation a computationally challenging task. 
Owing to symmetry the $PO$ region is always centered on $r_0=0$, and states with negative winding number are found in regions obtained by reflection in $r_0=0$.

\begin{figure}
{}\hspace{.5cm}(a) $a=1.0000$ \hspace{2.5cm}(b)  $a=1.0025$\hspace{2.5cm}(c) $a=1.0050$\\
\includegraphics[width=50mm]{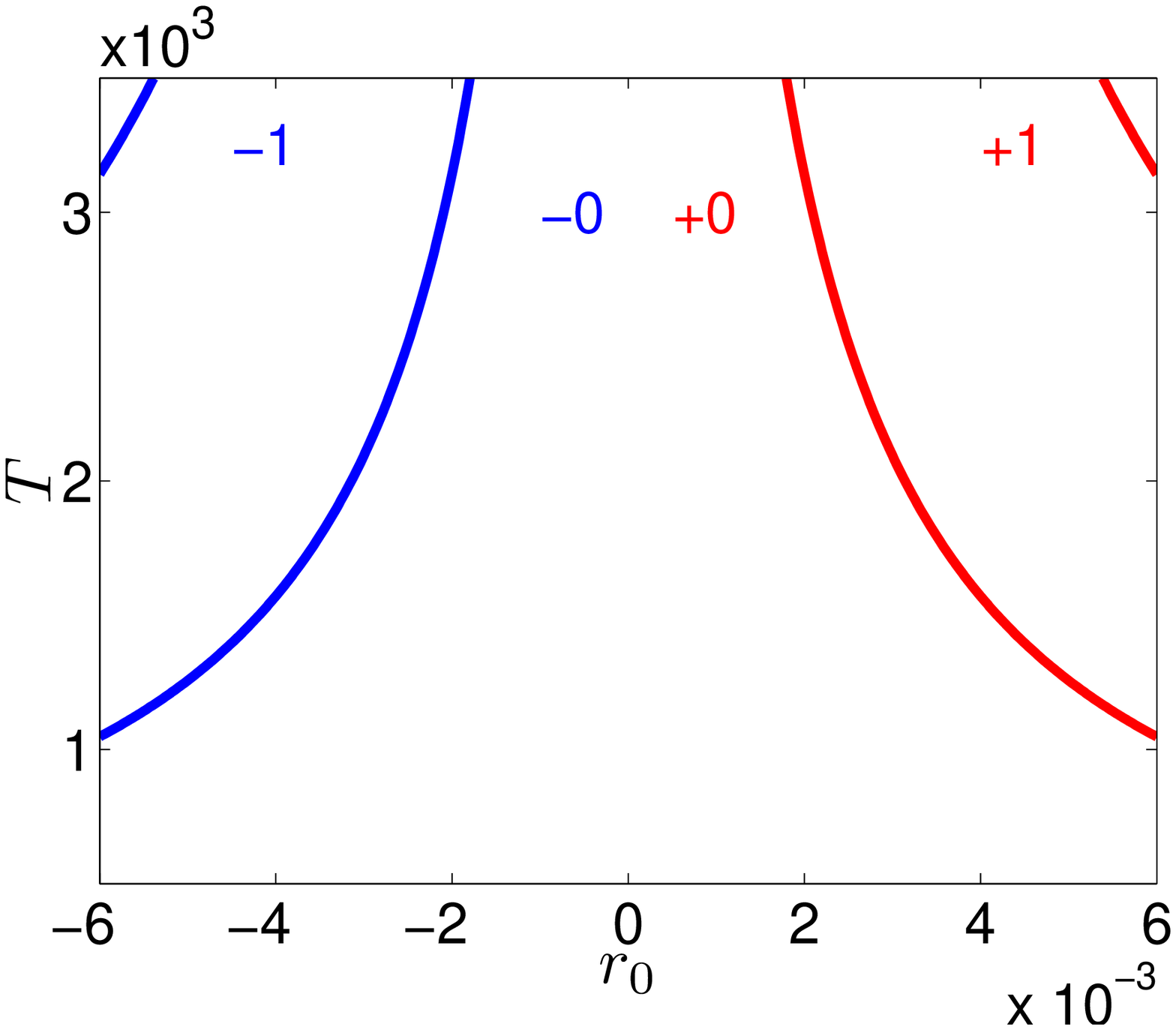}
\includegraphics[width=50mm]{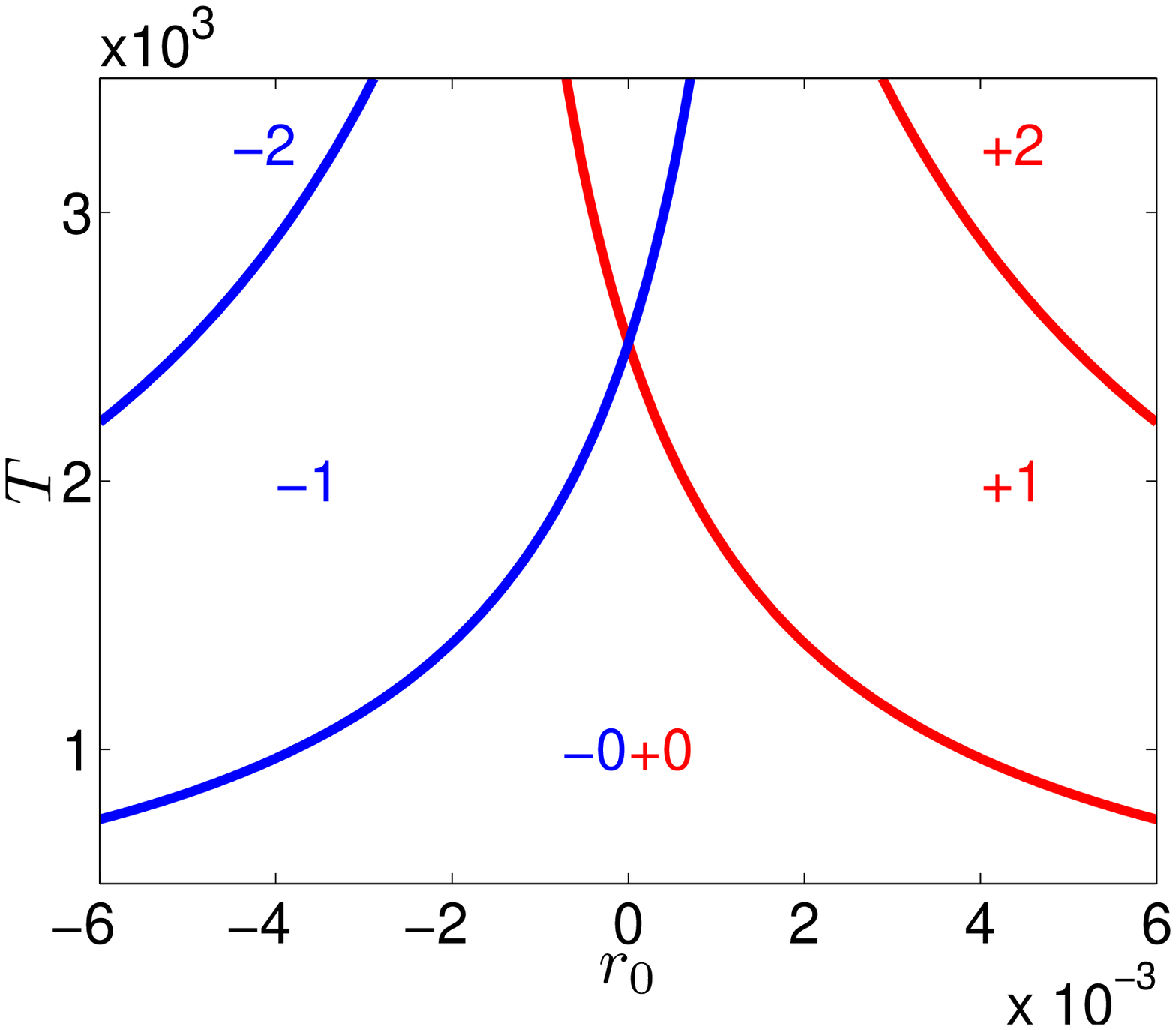}
\includegraphics[width=50mm]{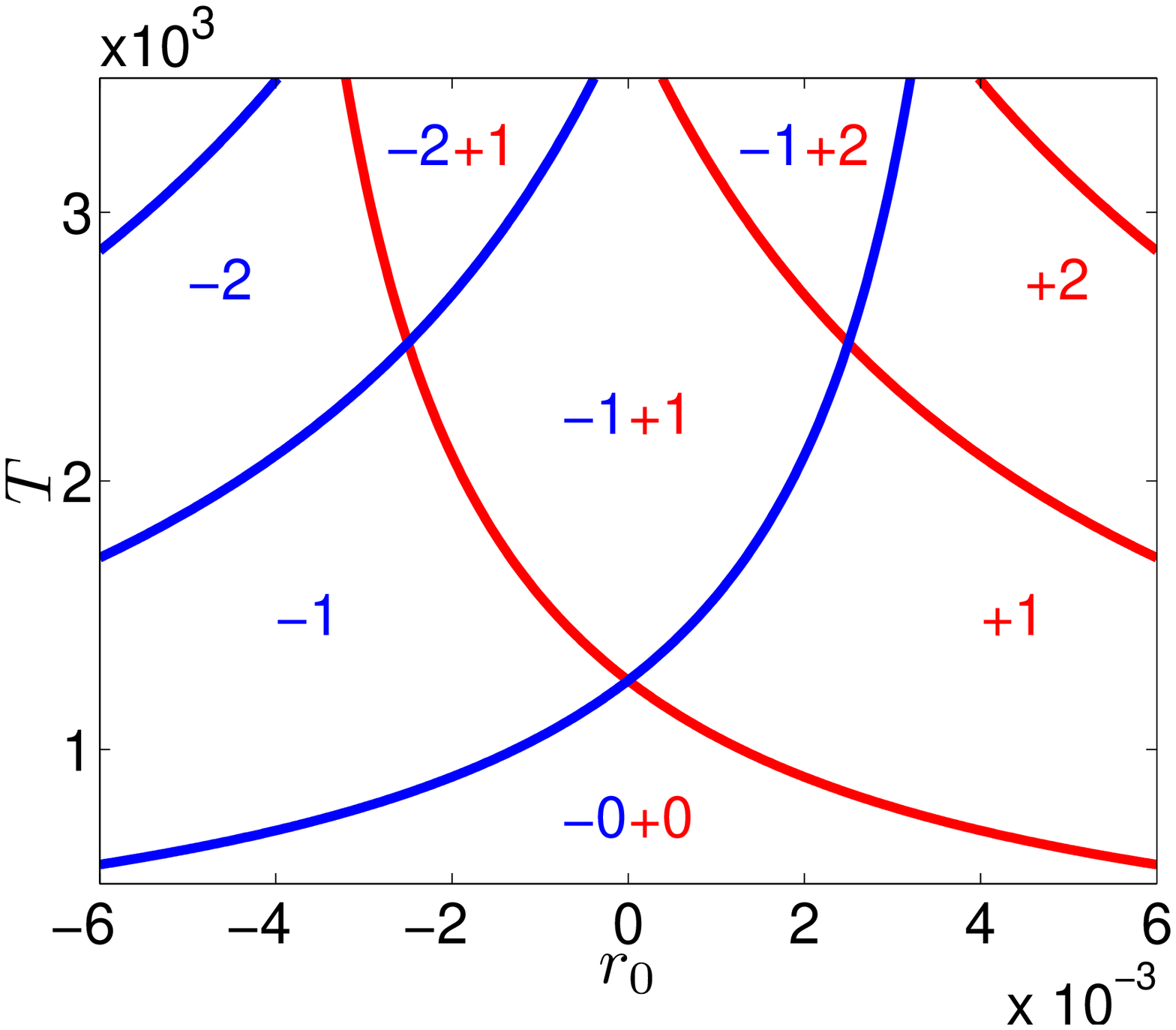}
\caption{(Color online) Transitions between the regions of constant $n_+$ (red, negative slope) and $n_-$ (blue, positive slope) in the $(r_0,T)$ plane as predicted by the asymptotic theory (Eq.~(\ref{eq:asymptoticN})) for $a=1.0000,\;1.0025,\; 1.0050$, respectively.  A sweet spot and pinching structure begins to emerge as $a$ increases. }
\label{fig:asympSS}
\end{figure}

The figure reveals the formation of sweet spots in this limit whenever $a>1$. When $a<1$, there are two distinct sets of resonance bands -- one set formed by regions with a fixed number of positive phase slips $n_+$, and the other by regions with a fixed number of negative phase slips $n_-$.  At $a=1$ the two sets of resonance bands both asymptote to $r_0=0$ as $T\rightarrow \infty$ (Fig.~\ref{fig:asympSS}(a)). The sweet spots and pinched zones emerge through the intersections of these resonance bands that take place once $a>1$ (Fig.~\ref{fig:asympSS}(b,c)). In particular, the pinched zone separating the $n$ and $n+1$ sweet spots in the $PO$ region is located at $(a-1)T/4\pi=n+1/2$ and  marks the transition from $n$ to $n+1$ positive and negative phase slips within a modulation cycle.       

\section{Adiabatic Theory}
\label{sec:adiabatic}

We now consider a more general time-dependence for the parameter $r$, but assume it varies slowly enough that we can treat the dynamics quasi-statically: $r=r(2\pi t/T)$ with $T\gg 1$. In this adiabatic limit, two distinct types of dynamics arise: slow dynamics that track the steady state phase-locked solution when $-1 \lesssim r( 2\pi t/T) \lesssim 1$ and a fast phase rotation with an adiabatically varying parameter when $|r(2\pi t/T)|\gtrsim 1$. No matter how low the frequency is, there is always an intermediate regime around the transition from a phase-locked state to rotation where the phase rotation is slow enough that it occurs on the same scale as the parameter drift. We apply WKB theory to capture the dynamics in each of the two regions separately and provide a condition for matching the solution across the transitions at $r\approx \pm 1$. 

We start with Eq.~(\ref{eq:hill}) but assume that $r=r(\omega t)$ with $\omega\ll 1$. We do not need to specify the form of $r(\omega t)$.
We transform this equation into a standard form for WKB theory by recasting it in terms of the slow time $\phi=\omega t$:
\begin{equation}
z'' +\frac{1}{\omega^2}\left(\frac{r^2-1}{4}+\frac{\omega r '}{2r}+\frac{\omega^2 r''}{2 r}-\frac{3\omega^2 (r')^2}{4r^2}\right)z=0.
\end{equation}
The system transitions from a phase-locked state to winding near $r^2-1\sim\mathcal{O}({\omega})$, and we can use the standard WKB ansatz $z=A z_\mathrm{WKB}= A \exp(iS/\omega) + \rm{c.c.}$, where $A$ is an arbitrary complex constant determined from initial conditions and/or matching procedures when we are away from these points. We suppose that $S=S_0+\omega S_1+\dots$ and match orders to solve for each $S_i$.  

Making the WKB substitution generates, at leading order,
\begin{equation}
S_0'^2=\frac{r^2-1}{4}.
\end{equation} 
The leading order WKB solution, in terms of the original time scale, is $z=A \exp\left(\pm \tfrac{i}{2}\int \sqrt{r^2-1} dt\right)$.  The equation at next order is, after simplification,
\begin{equation}
-2S_0' S_1'+i S_0''+\frac{r'}{2r}=0,
\end{equation}
yielding
\begin{equation}
S_1=\frac{1}{2} \left( i  \log \sqrt{r^2-1}\mp\tan ^{-1}\left(\frac{1}{\sqrt{r^2-1}}\right)\right),
\end{equation}
depending on the choice of root for $S_0$. 

Including this correction, the solution becomes
\begin{equation}
\label{zoutpl}
z=\frac{A}{(r^2-1)^{1/4}}\exp\pm\tfrac{i}{2}\left(\int \sqrt{r^2-1} dt- \tan^{-1}\frac{1}{\sqrt{r^2-1}} \right).
\end{equation}
When $r<1$, we find it convenient to rewrite the expression for $S_1$ as
\begin{equation}
S_1=\frac{i}{2} \left( \log \sqrt{1-r^2}\pm \log\frac{1+\sqrt{1-r^2}}{r}\right),
\end{equation}
and the solution now takes the form
\begin{equation}
\label{zinpl}
z=\frac{A}{(1-r^2)^{1/4}}\left(\frac{1+\sqrt{1-r^2}}{r}\right)^{\mp 1/2}\exp\mp\tfrac{1}{2}\int \sqrt{1-r^2} dt.
\end{equation}

Near a transition point $r=1$ (we take it to be at $t=0$), we suppose that $r\approx 1+\alpha t$, where $\alpha=\dot{r}(0) =\omega r'(0)$ is a constant.  To leading order, the equation becomes
\begin{equation}
\label{airyfuncsol}
\ddot{z}+\frac{\alpha}{2} (t+1) z=0,
\end{equation}
which has solutions in terms of the Airy functions $\mathrm{Ai}(s)$ and $\mathrm{Bi}(s)$, where $s=-\left(\frac{\alpha}{2}\right)^{1/3}(t+1)$.

We will further assume $\alpha>0$ so that the transition occurs as the system leaves the phase-locked region and enters the winding region, and remind the reader that
\begin{equation}
\tan\frac{\theta}{2} = -\frac{2}{r}\frac{\dot{z}}{z}+\frac{1}{r}\left(1-\frac{\dot{r}}{r}\right).
\end{equation}

\begin{figure}
\includegraphics[width=150mm]{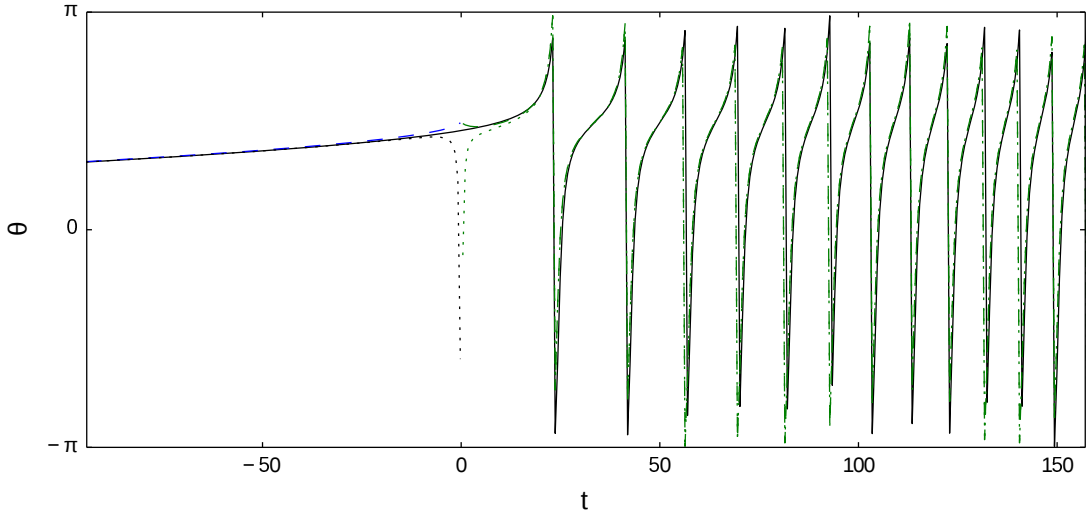}
\caption{(Color online) The phase $\theta(t)$ $\mathrm{mod} (2\pi)$ near the transition from a phase-locked state to winding from a time simulation with $T=2\pi\times10^{3}$, $a=2$, and $r_0=0$ (black solid line). The simulation represents the evolution of $\theta$ in the time window $[-100;160]$ of a converged periodic orbit for which $t=0$ corresponds to $r=1$. The dashed lines are computed using the adiabatic predictions (\ref{adiabthpl}), (\ref{adiabthw}) without the ``subdominant'' term proportional to $\dot{r}$ while the dotted lines take it into account. Predictions in the phase-locked (winding) regime are shown in blue (green) for $t<0$ ($t>0$).}
\label{fig:checkwkb}
\end{figure}

We consider the solution within the phase-locked region that follows, for $t<0$, the stable steady-state solution branch $\theta=\sin^{-1}r$, corresponding to taking the negative root of $S_0$. Thus, equation (\ref{zinpl}) reduces to
\begin{equation}
z=\frac{A_\mathrm{pl}}{(1-r^2)^{1/4}}\left(\frac{1+\sqrt{1-r^2}}{r}\right)^{ 1/2}\exp\tfrac{1}{2}\int \sqrt{1-r^2} dt,  \label{eq:zpl}
\end{equation}
where $A_\mathrm{pl}$ depends on the choice of initial condition.
In terms of $\theta$, expression~\ref{eq:zpl} reads
\begin{equation}
\label{adiabthpl}
\tan \frac{\theta}{2}=\frac{1-\sqrt{1-r^2}}{r}\left(1-\frac{\dot{r}}{r(1-r^2)}\right).
\end{equation}
In order to match solutions across the transition region, we must take the $t\rightarrow 0$ (equivalently $r \rightarrow 1$ limit of this solution and match it to the $t\rightarrow -\infty$ (equivalently $s \rightarrow \infty$) limit of the Airy function solution of Equation (\ref{airyfuncsol}).
This procedure selects the Airy function Ai with amplitude proportional to $A_\mathrm{pl}$.
On the other side, the winding solution coming from equation (\ref{zoutpl}) can be matched to the Airy solution when written in the form 
\begin{equation}
\label{zmatchw}
z=\frac{A_\mathrm{w}}{(r^2-1)^{1/4}}\cos\frac{1}{2}\left(\int \sqrt{r^2-1} dt-\tan^{-1}\frac{1}{\sqrt{r^2-1}}\right),
\end{equation}
where $A_\mathrm{w} = A_\mathrm{w} (A_\mathrm{pl})$.
The matching is achieved by comparing the $r \rightarrow 1$ ($t\rightarrow 0$) limit of expression (\ref{zmatchw}) to the $t \rightarrow \infty$ ($s \rightarrow - \infty$) limit of the Airy function obtained in the matching procedure with the phase-locked solutions.
Expression (\ref{zmatchw}) yields:
\begin{equation}
\label{adiabthw}
\tan \frac{\theta}{2}=\frac{1}{r}\left[1+\sqrt{r^2-1}\tan\frac{1}{2}\left(\int \sqrt{r^2-1} dt-\tan^{-1}\frac{1}{\sqrt{r^2-1}}\right)\right]\left(1-\frac{\dot{r}}{r(r^2-1)}\right).
\end{equation}
Figure~\ref{fig:checkwkb} shows a comparison of the WKB solution in terms of $\theta$ with a periodic orbit obtained through simulation with $r(t)=2\sin (10^{-3} t+\pi/6)$.

The results obtained from the WKB approximation in the limit of a slowly-varying frequency parameter can be generalized using a theorem that places bounds on the number of zeros of solutions to linear second order differential equations. Given an equation of the form
\begin{equation}
\ddot{z}+q(t)z=0
\end{equation}
with $q(t)>0$ in $C^2$ and bounded, such that $\dot{q}(t)=o\left(q^{3/2}(t)\right)$ as $t\rightarrow \infty$, it can be shown~\cite{hartman1982ode} that the number of zeros $[n]$ between $0\le t\le T$ for a given solution $z(t)\neq 0$ is bounded by
\begin{equation}
\left\lvert \pi [n]-\int_0^T \sqrt{q(t)}dt\right\rvert\leq \pi +\int_0^T\left\lvert \frac{5\dot{q}^2}{16q^{5/2}}-\frac{\ddot{q}}{4q^{3/2}} \right\rvert dt.
\label{eq:generalization}
\end{equation}
 It follows that when $\dot{q}\ll 1$
\begin{equation}
\pi [n] \sim \int_0^T \sqrt{q(t)}dt \qquad \mathrm{as}\quad T\rightarrow\infty,
\label{eq:bound}
\end{equation}
thereby reproducing the quasi-static prediction from WKB theory. In the case of the Adler equation, the corresponding frequency parameter is given by
\begin{equation}
q(t)=\frac{r^2-1}{4} + \frac{\dot{r}}{2r}+\frac{\ddot{r}}{2r}-\frac{3\dot{r}^2}{4r^2}.
\end{equation}
The conditions on $r$ for the applicability of the bound within the time interval of interest are that $|q(t)|>0$, $r\in C^4$ and is bounded. We can make some further approximations in the limit that $r=r(\omega t)$ is slowly varying, i.e., $\omega\ll1$, and the first condition reduces to $|r|+\mathcal{O}(\omega)>1$.  In this adiabatic limit, the integral in the bound becomes
\begin{equation}
\int \sqrt{q(t)}dt= \int \frac{\sqrt{r^2-1}}{2}dt-\tan^{-1}\frac{1}{\sqrt{r^2-1}}+\mathcal{O}(\omega).
\end{equation}

The bound on the number of zeros of the Hill equation translates into a bound on the number of phase slips incurred by a solution to the Adler equation over a given time interval where $q(t)>0$, i.e., when $r(t)$ is outside of the phase-locking region.  We define $n_\pm$ by the integral
\begin{equation}
n_\pm=\frac{1}{\pi}\int_{\mathcal{T}_\pm} \sqrt{q(t)}dt
\end{equation}
over the time interval $\mathcal{T}_\pm$ spent with $q(t)>0$ and $r(t)>1$ ($r(t)<-1$) for $n_+$ ($n_-$).  The bound described above restricts the number of phase slips over $\mathcal{T}_\pm$ to either rounding up or down ($\lfloor n_\pm \rfloor$ or $\lceil n_\pm \rceil$) to order $\mathcal{O}(\omega)$. This is a generalization of the WKB solution in the sense that the bound applies even when the slowly-varying assumption does not hold.  Some care must be taken when applying this bound as $q\to \infty$ as $r\to 0$.  The bound must be applied to positive and negative phase slips separately in order to place a bound on the winding number of a particular trajectory.  

\begin{figure}
{}\hspace{1cm}(a) WKB  \hspace{6cm} (b) bound \\
\includegraphics[width=75mm]{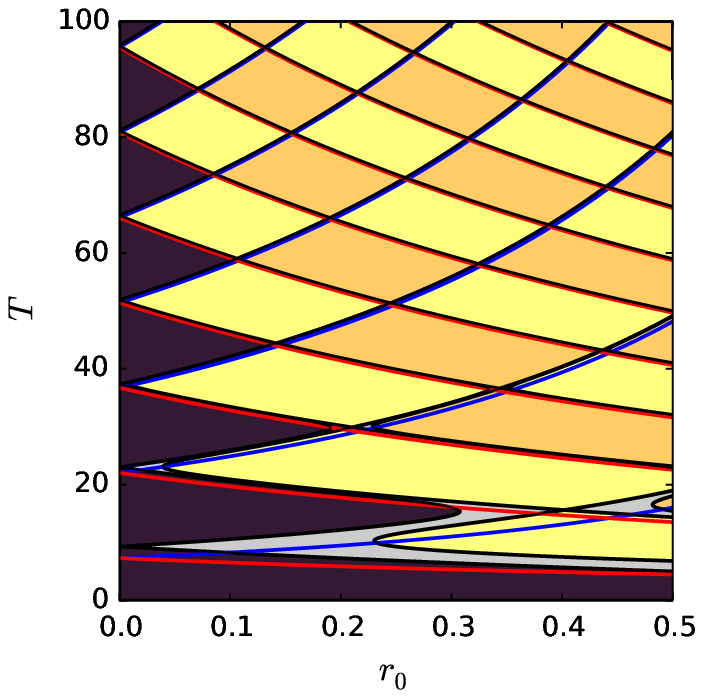}
\includegraphics[width=75mm]{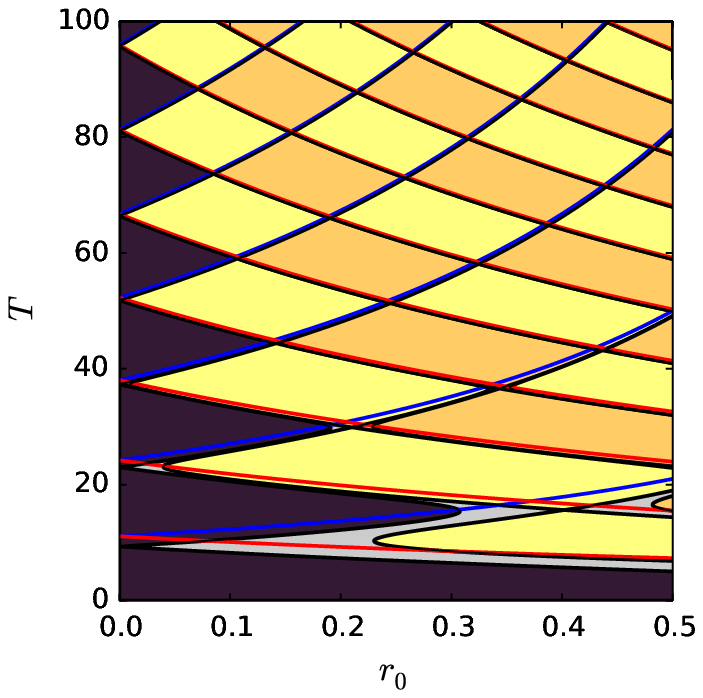}
\caption{(Color online) Average winding number per period $T$ of the frequency parameter shown in the $(r_0,T)$ plane for $a=2$. Colors represent results from numerical simulation: no net phase slips occur over the course of a modulation period in the dark region to the left; the alternating lighter yellow/darker orange regions to the right indicate $1,2,3,\dots$ net phase slips as $r_0$ increases. The red/blue (negative/positive slope) lines represent predictions of adiabatic theory. The left panel shows the prediction based on the WKB approximation (Eq.~(\ref{eq:wkb})) while the right panel shows the prediction based on the bound in Eq.~(\ref{eq:bound}).}
\label{fig:adiabatic}
\end{figure}

The WKB approximation can be used to predict the partitioning of the parameter space by winding number (see Fig.~\ref{fig:rTplane}) by computing the net winding number $N=[n_+]+[n_-]$, where 
\begin{equation}
n_\pm =\pm\frac{T}{2\pi^2}\int_{\phi_\pm}^{\pi/2}\sqrt{(r_0\pm a\sin\phi)^2-1}\,d\phi,
\label{eq:wkb}
\end{equation}
and $r_0 \pm a\sin\phi_\pm=1$. The first correction from WKB theory cancels because the system always enters and exits the phase-locked region at the same value of $r$.  Replacing the expression in the square root with $q(t)$ provides a way to estimate the winding number from the bound. Figure~\ref{fig:adiabatic} shows a comparison of the resulting prediction with the numerical results in Fig.~\ref{fig:rTplane}.  We see that the adiabatic theory agrees well with the numerical results far beyond the low frequency limit for which it was constructed, a conclusion supported by the generalization (\ref{eq:generalization}).

\section{Discussion}
\label{sec:concl}

In this paper, we have investigated the dynamics of two coupled oscillators when the frequency difference is modulated in time. The same equation describes a multitude of other systems, ranging from Josephson junctions to systems of large numbers of coupled oscillators as detailed in Sec.~\ref{sec:intro}. Specifically, we studied here the Adler equation \cite{adler1946study} with a sinusoidally varying frequency parameter. The frequency modulation introduces two new parameters into the problem, in addition to the mean frequency difference $r_0$: the amplitude $a$ and the period $T$ of the modulation. While the autonomous Adler equation leads to phase locking for $-1\le r_0 \le 1$ and persistent drift for $|r_0| > 1$, we have unveiled much richer dynamics that take place when frequency modulation is activated: the phase-locked solutions turn into periodic orbits and the phase difference $\theta$ between the oscillators becomes a periodic function of time. The region $PO$ of the existence of these periodic orbits is centered around $r_0 = 0$ and exhibits a succession of sweet spots as $a$ or $T$ increases, interspersed with pinched zones where the width of the $PO$ region vanishes. The width of these sweet spots decreases with increasing $a$ and $T$. On either side of $PO$ are regions within which the solution grows or decays by one, two, etc. phase slips per modulation cycle. These regions have the same basic structure as the $PO$ region and are separated by exponentially thin transition zones where the number of phase slips fluctuates from cycle to cycle. This intricate behavior is a consequence of a sequence of resonances between the time needed for a phase slip and the period of the modulation, and can be described, in an appropriate regime, in terms of an interaction between $n$:1 and $-n$:1 resonance tongues.

\begin{figure}
\includegraphics[width=75mm]{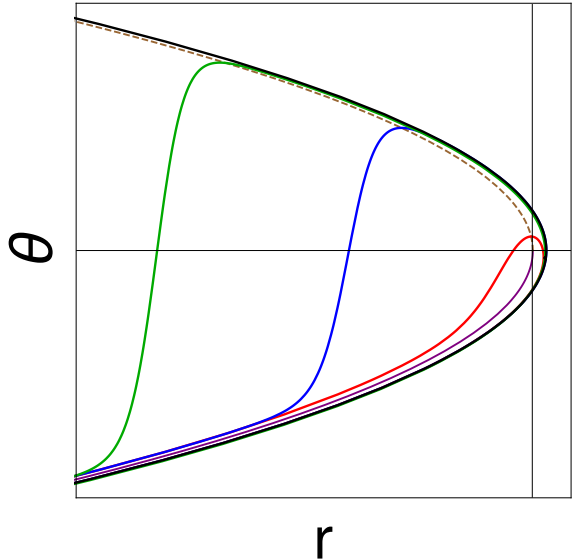}
\caption{(Color online) Canard behavior near $r=1$ in the limit $T\gg 1$ as predicted by Eq.~(\ref{eq:weber}) for $\nu = -10^{-1}$ (red, inner), $-10^{-6}$ (blue, middle), $-10^{-12}$ (green, outer) and $0$ (black). In terms of the parameters of the original problem $\nu\equiv\frac{1}{4\pi T}(r_0+a-1) - \frac{1}{2}$; the horizontal and vertical scales are $r-1\sim 1/T$ and $\theta-\pi/2 \sim 1/\sqrt{T}$. The stable (solid purple) and unstable (dashed brown) stationary solutions to the autonomous problem are shown for reference.    }
\label{fig:Canard}
\end{figure}

Canard orbits form an essential part of this picture~\cite{Krasnosel'skii2013periodic}. These are present in the vicinity of the boundaries of the $PO$ region and consist of trajectories that drift along a branch of stable equilibria for part of the cycle; after reaching a fold at which the equilibria lose stability the trajectory drifts for a time along the branch of unstable equilibria, instead of detaching, before an abrupt jump back to a stable equilibrium. Equation (\ref{eq:weber}) describes the emergence of such trajectories for low frequency modulation with mean near $r_0 = 0$ and amplitude slightly larger than $a=1$; Fig.~\ref{fig:Canard} shows several examples of the predicted canard solutions, for comparison with the ``larger'' periodic canard orbits computed numerically in Sec. \ref{sec:canard}.

\begin{figure}
\includegraphics[width=75mm]{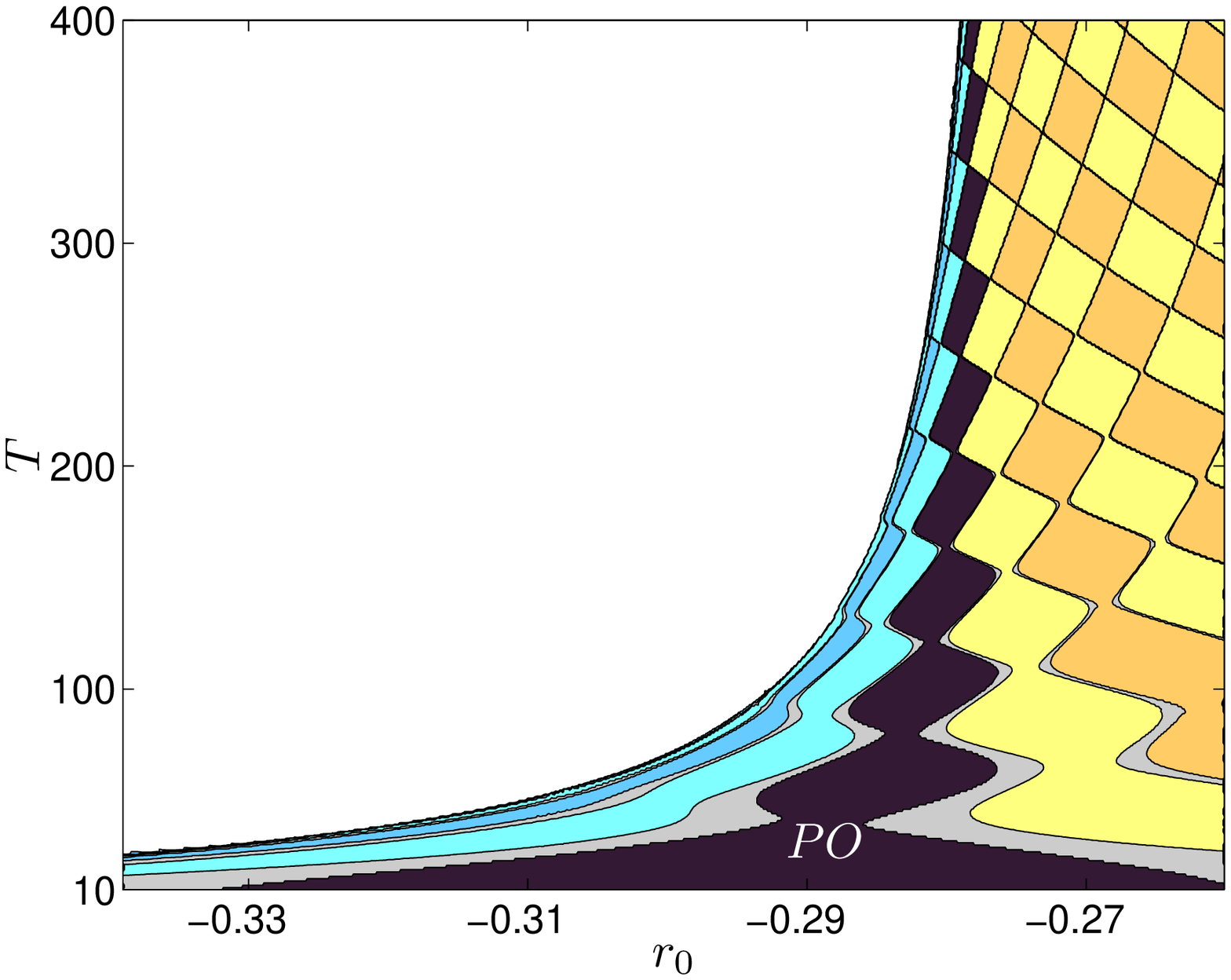}
\caption{(Color online) Average winding number per period $T$ shown in the $(r_0,T)$ plane for the Swift--Hohenberg equation when $b=1.8$ \cite{gandhi2015localized,gandhi2014new}. In the alternating lighter yellow and darker orange regions on the right, a localized state grows by the net addition of $2,4,6,\dots$ wavelengths of the pattern within each forcing cycle, while in the alternating light and dark blue regions on the left it shrinks by $2,4,6,\dots$ wavelengths within each cycle; the dark region labeled $PO$ corresponds to localized states that pulsate but maintain constant average length. The gray areas represent transition zones where the average winding number per period is not an integer. In this system, such zones are characterized by a devil's staircase type structure.}
\label{fig:rTplaneSHE}
\end{figure}
We mention that similar behavior has been observed in the partial differential equation description of the dynamics of spatially localized states \cite{gandhi2015localized}. In this work, the quadratic-cubic Swift--Hohenberg equation (SHE23) is forced in a time-periodic manner and a similar partitioning of parameter space is observed (Fig. \ref{fig:rTplaneSHE}). The reason for this similarity can be traced to the nature of the motion, under parametric forcing, of fronts connecting a spatially periodic state of SHE23 to the trivial, homogeneous state: the front motion is analogous to repeated phase slips, with each ``phase slip'' corresponding to a nucleation or annihilation event that adds or subtracts one wavelength of the pattern at either end of the localized structure. However, the resulting partitioning of the parameter space is not symmetric owing to a lack of symmetry between positive and negative ``phase slips''. An adiabatic theory of the type described here works equally well in SHE23 and its predictions are in excellent agreement with the results of numerical simulations \cite{gandhi2014new}. Indeed SHE23 also displays canards associated with the transitions between different states (lightest gray regions in Fig. \ref{fig:rTplaneSHE} \cite{Gandhi2016}).

The work presented here has a direct application to Josephson junctions driven by an AC current. In the overdamped limit such junctions are modeled by Eq.~(\ref{eq:adler})~\cite{likharev1986dynamics}, with $r$ representing the external drive and $\theta$ the phase difference of the Ginzburg--Landau order parameter across the gap. The so-called supercurrent across the gap is proportional to $\sin\theta$ while the voltage produced corresponds to the derivative $\dot{\theta}$. In this context, phase-locking and phase-slips are closely related to the existence of Shapiro steps~\cite{shapiro1967microwave} for a single Josephson junction. Related dynamics arise in arrays of Josephson junctions that are globally coupled via an LRC circuit~\cite{wiesenfeld1998frequency}. These systems provide a physical realization of the phase-coupled oscillator models mentioned in the introduction.  

In fact, weakly coupled systems can often be decomposed into two parts with part A obeying dynamics that are largely insensitive to the dynamics of part B. In these circumstances it often suffices to consider system B on its own but with prescribed time-dependence arising from the coupling to A. This is the case, for example, in globally coupled phase oscillator systems, in which each oscillator responds to the global dynamics of the system but the global dynamics are insensitive to the details of the dynamics of an individual oscillator. These systems, for reasons explained in the introduction, have properties closely related to the nonautomous Adler equation studied here. For these reasons we anticipate applications of the techniques developed here to studies of synchronization in oscillator networks.  

\begin{acknowledgments}
This work was supported in part by the National Science Foundation under grants DMS-1211953 and CMMI-1233692.  We thank Benjamin Ponedel for insightful discussions.
\end{acknowledgments}

\bibliography{AdlerBib}

\end{document}